%% file: ZprimeTauTau2012Paper.tex
\pdfoutput=1
\documentclass[11pt,a4paper]{article} 
\usepackage{jheppub}
\usepackage{atlasphysics}
\usepackage{authoraftertitle}
\usepackage{amsmath}
\usepackage{graphicx}
\DeclareGraphicsExtensions{.pdf}
\usepackage{booktabs} 
\usepackage{siunitx}
\usepackage{aliases}
\usepackage{placeins} 
\usepackage{preprintcover}

\makeatletter
\g@addto@macro\bfseries{\boldmath}
\makeatother

\title{A search for high-mass resonances decaying to $\tau^{+}\tau^{-}$ in $pp$ collisions at $\sqrt{s}=8\TeV$ with the ATLAS detector}

\collaboration{The ATLAS Collaboration}

\abstract{
A search for high-mass resonances decaying into $\tau^{+}\tau^{-}$ final states
using proton--proton collisions at $\sqrt{s}= 8$~TeV produced by the Large
Hadron Collider is presented. The data were recorded with the ATLAS detector
and correspond to an integrated luminosity of 19.5--20.3~fb$^{-1}$.  No
statistically significant excess above the Standard Model expectation is
observed; 95\% credibility upper limits are set on the cross section times
branching fraction of $Z^{\prime}$ resonances decaying into $\tau^+\tau^-$
pairs as a function of the resonance mass.  As a result, $Z^{\prime}$ bosons of
the Sequential Standard Model with masses less than 2.02~TeV are excluded at
95\% credibility. The impact of the fermionic couplings on the $Z^{\prime}$
acceptance is investigated and limits are also placed on a $Z^{\prime}$ model
that exhibits enhanced couplings to third-generation fermions. 
}

\PreprintCoverPaperTitle{A search for high-mass resonances decaying to $\tau^{+}\tau^{-}$ in $pp$ collisions at $\sqrt{s}=8\TeV$ with the ATLAS detector} 

\PreprintIdNumber{CERN-PH-EP-2015-018}

\PreprintCoverAbstract{
A search for high-mass resonances decaying into $\tau^{+}\tau^{-}$ final states
using proton--proton collisions at $\sqrt{s}= 8$~TeV produced by the Large
Hadron Collider is presented. The data were recorded with the ATLAS detector
and correspond to an integrated luminosity of 19.5--20.3~fb$^{-1}$.  No
statistically significant excess above the Standard Model expectation is
observed; 95\% credibility upper limits are set on the cross section times
branching fraction of $Z^{\prime}$ resonances decaying into $\tau^+\tau^-$
pairs as a function of the resonance mass.  As a result, $Z^{\prime}$ bosons of
the Sequential Standard Model with masses less than 2.02~TeV are excluded at
95\% credibility. The impact of the fermionic couplings on the $Z^{\prime}$
acceptance is investigated and limits are also placed on a $Z^{\prime}$ model
that exhibits enhanced couplings to third-generation fermions. 
}

\PreprintJournalName{JHEP}

\begin{document}

\maketitle
\flushbottom

\input{intro}

\input{experiment}

\input{samples}

\input{reco}

\input{sel}

\input{bkg}

\input{sys}

\input{models}

\input{results}

\FloatBarrier
\input{summary}

\input{Acknowledgements}

\bibliographystyle{JHEP}
\bibliography{ZprimeTauTau2012Paper}

\onecolumn
\clearpage 
\input{atlas_authlist}

\end{document}

%% file: intro.tex
\section{Introduction}
\label{sec:introduction}

Searches for new heavy resonances decaying to tau lepton pairs are both
theoretically and experimentally well
motivated~\cite{Hewett:1989,Cvetic:1995zs,Leike:1998wr,Rizzo:2006nw,Diener:2010sy,Langacker:2008}.
Heavy \Zprime{} bosons often arise in grand unified theories and while they are
typically considered to obey lepton universality, this is not necessarily a
requirement. In particular, some models offering an explanation for the high
mass of the top quark predict that such bosons preferentially couple to
third-generation fermions~\cite{ZprimeToThirdGen,SFM}.  Models containing
non-universal \Zprime{} bosons can explain the anomalous dimuon production
observed at the D0 experiment~\cite{DimuonExcess,DimuonD0} and the excess in
semileptonic $B$-meson decays into tau leptons observed at the Belle and BaBar
experiments~\cite{NonUniversalLR,BDtaunuBelle,BDtaunuBaBar}.  Searches in the
\ditau{} channel are also sensitive to sgoldstino-like scalars in
supersymmetric models~\cite{LowScaleSusyDitau,sgoldstinos}, hidden sector
\Zprime{} models~\cite{HiddenSectorZprimes} and to the anomalous tau lepton
dipole moments and higher-order tau--gluon couplings~\cite{TauGluonCouplings}.

In this article, a search for high-mass resonances decaying into \tptm{} final
states using proton-proton ($pp$) collisions at a center-of-mass energy of
$\sqrt{s}= 8\TeV$ produced by the Large Hadron Collider (LHC)~\cite{LHC:2008}
is presented. The data were recorded with the ATLAS
detector~\cite{AtlasDetectorPaper} and correspond to an integrated luminosity
of 19.5--20.3~fb$^{-1}$. Tau leptons can decay into a charged lepton and two
neutrinos ($\taul = \taue$ or \taumu{}), or hadronically (\tauh{}),
predominantly into one or three charged pions, a neutrino and often additional
neutral pions.  The \hadhad{}, \muhad{} and \ehad{} channels are analysed,
accounting for 42\%, 23\% and 23\% of the total \tptm{} branching fraction,
respectively.  A counting experiment is performed in each channel from events
that pass a high-transverse-mass requirement.  Due to the different dominant
background contributions and signal sensitivities, each channel is analysed
separately and a statistical combination is used to maximise the sensitivity.  

The Sequential Standard Model (SSM), which contains a \ZprimeSSM{} boson with
couplings identical to the Standard Model $Z$ boson, is chosen as the benchmark
model to optimise the analysis and to quantify the experimental sensitivity.
Limits on the \ZprimeSSM{} cross section times the branching fraction in tau pairs,
\mbox{$\sigma(pp\to\ZprimeSSM+X)\cdot\mathcal{B}(\ZprimeSSM\to\tau^+\tau^-)\equiv\xsecbrssm$},
are provided as a function of the resonance mass, \mzp{}.  The impact on the
signal acceptance times efficiency from changing the \ZprimeSSM{} couplings is
assessed, which allows the limits on \ZprimeSSM{} to be reinterpreted for a
broad range of models.  Limits are also placed on the non-universal $G(221)$
model~\cite{SFM,Topflavor,G221}, which contains a \ZprimeSFM{} boson that can
exhibit enhanced couplings to tau leptons. 

Direct searches for high-mass \ditau{} resonances have been performed by the
ATLAS and CMS collaborations using 5~\ifb{} of integrated luminosity at
$\sqrt{s}=7\TeV$~\cite{ATLASZprime5fb,CMSZprime5fb}. The searches exclude
\ZprimeSSM{} with masses below $1.4\TeV$ at 95\%~CL.\footnote{CL is used
interchangeably throughout this article to refer to both confidence level
(frequentist) and credibility limit (Bayesian).} For comparison, the most
stringent limits on \ZprimeSSM{} in the \diele{} and \dimuon{} decay channels
combined are $2.90\TeV$ at 95\%~CL from both ATLAS~\cite{ATLASdilepton} and
CMS~\cite{CMSdilepton8TeVpaper}.  While the limits on
\mbox{$\sigma(pp\to\ZprimeSSM+X)\cdot\mathcal{B}(\Zprime\to e^+e^-/\mu^+\mu^-)$} are
in general stronger than those on \xsecbrssm{}, they may be evaded by models
with weak couplings to electrons and muons.  Indirect limits on \Zprime{}
bosons with non-universal flavour couplings have been set using measurements
from LEP and LEP~II~\cite{ZprimeMassLimitsFromLEP} and translate to a lower
bound on the \Zprime{} mass of $1.09\TeV$ at 95\%~CL.  Indirect limits have
also been placed on the non-universal $G(221)$
model~\cite{G221LHC,LeeLFV,LeeCKM,SFM}. The strongest exclude \ZprimeSFM{} with
a mass lower than $1.8\TeV$ at 95\%~CL. 

This article is structured as follows. Section~\ref{sec:experiment} provides an
overview of the ATLAS detector.  The event samples used in the analysis,
recorded by the ATLAS detector or simulated using the ATLAS simulation
framework, are described in section~\ref{sec:samples}.  The reconstruction of
physics objects within the event samples is described in
section~\ref{sec:reco}. A description of the selection criteria used to define
\Zprime{} signal regions is given in section~\ref{sec:sel}.
Section~\ref{sec:bkg} describes the estimation of background contributions,
followed by a description of systematic uncertainties in section~\ref{sec:sys}.
In section~\ref{sec:models}, the impact of altering the \Zprime{} couplings on
the signal acceptance is described and the non-universal $G(221)$ model is
introduced.  A presentation of the results is given in
section~\ref{sec:results}, followed by concluding remarks in
section~\ref{sec:summary}.

%% file: experiment.tex
\section{ATLAS detector}
\label{sec:experiment}

The ATLAS detector at the LHC covers nearly the entire solid angle around the
collision point. It consists of an inner tracking detector surrounded by a thin
superconducting solenoid, electromagnetic (EM) and hadronic calorimeters, and a
muon spectrometer incorporating large superconducting toroid magnets.

The inner-detector system is immersed in a 2~T axial magnetic field and
provides charged-particle tracking in the range $|\eta| < 2.5$.\footnote{ATLAS
uses a right-handed coordinate system with its origin at the nominal
interaction point (IP) in the centre of the detector and the $z$-axis along the
beam pipe. The $x$-axis points from the IP to the centre of the LHC ring, and
the $y$-axis points upward. Cylindrical coordinates $(r,\phi)$ are used in the
transverse plane, $\phi$ being the azimuthal angle around the beam pipe. The
pseudorapidity is defined in terms of the polar angle $\theta$ as
$\eta=-\ln\tan(\theta/2)$. The geometrical distance between objects is defined
as $\Delta R=\sqrt{(\Delta\phi)^2 + (\Delta\eta)^2}$. } A high-granularity
silicon pixel detector covers the vertex region and typically provides three
measurements per track. It is followed by a silicon microstrip tracker, which
usually provides four pairs of measurements per track.  These silicon detectors
are complemented by a transition radiation tracker (TRT), which enables
radially extended track reconstruction up to $|\eta| = 2.0$. The TRT also
provides electron/pion discrimination based on the fraction of hits (typically
30 in total) above a higher energy-deposit threshold corresponding to
transition radiation.  

The calorimeter system covers the pseudorapidity range $|\eta|< 4.9$. Within
the region $|\eta|< 3.2$, EM calorimetry is provided by high-granularity
barrel and endcap liquid-argon (LAr) EM calorimeters with lead absorbers, with an additional
thin LAr presampler covering $|\eta| < 1.8$ to correct for upstream energy
loss. Hadronic calorimetry is provided by a steel/scintillator-tile
calorimeter, segmented into three barrel structures within $|\eta| < 1.7$, and
two copper/LAr hadronic endcap calorimeters. Coverage in the forward
region is achieved by copper/LAr and tungsten/LAr calorimeter modules
optimised for EM and hadronic measurements, respectively.  

The muon spectrometer comprises separate trigger and high-precision
tracking chambers measuring the deflection of muons in a magnetic field
generated by superconducting air-core toroids. The precision chamber system
covers the region $|\eta| < 2.7$ with three layers of monitored drift tubes,
complemented by cathode strip chambers in the forward region, where the
background is highest. The muon trigger system covers the range $|\eta| < 2.4$
with resistive plate chambers in the barrel, and thin gap chambers in the
endcap regions.  

A three-level trigger system is used to select interesting
events~\cite{ATLASTrigger}. The Level-1 trigger is implemented in hardware and
uses a subset of detector information to reduce the event rate to a design
value of at most 75~kHz. This is followed by two software-based trigger levels
which together reduce the event rate to a maximum of 1~kHz.

%% file: samples.tex
\section{Event samples}
\label{sec:samples}

The data used in this search were recorded with the ATLAS detector in $pp$
collisions at a centre-of-mass energy of $\sqrt{s}=8\TeV$ during the 2012 run
of the LHC.  Only data taken with $pp$ collisions in stable beam conditions and
with all ATLAS subsystems operational are used, resulting in an integrated
luminosity of 20.3\,\ifb{}. For the analysis of the \hadhad{} channel, a small
fraction of data from the initial running period are discarded as the
trigger conditions are not accounted for by the simulation, resulting in an
integrated luminosity of 19.5~\ifb{}.  
The \hadhad{} channel uses events passing a single-tau trigger with a
transverse momentum (\pt{}) threshold of $125\GeV$, designed to select hadronic
tau decays.  The \muhad{} channel uses events passing a single-muon
trigger, either with a \pt{} threshold of $24\GeV$ including an isolation requirement
or with a threshold of $36\GeV$ without an isolation requirement.  The $\ehad$
channel uses events passing a single-electron trigger, either with a \pt{}
threshold of $24\GeV$ including an isolation requirement, or with a threshold
of $60\GeV$ without an isolation requirement. Events that pass the trigger are
selected if they contain a vertex with at least four associated tracks, each
with \mbox{$\pt >0.5\GeV$}.  Events may have several vertices satisfying this
requirement due to multiple $pp$ interactions occurring in the same or 
neighbouring bunch crossings, referred to as pile-up.  The event vertex is
chosen as the one with the largest sum of the squared track transverse momenta.

\begin{sloppypar} 
Monte Carlo (MC) simulation is used to estimate signal efficiencies and some
background contributions.  Simulated samples of events from the following
background processes are used: \DY{} and \DYlljets{} ($\ell=e,\mu$) enriched in
high-mass events, and \Wjets{}, \ttbar{}, single-top-quark ($Wt$, $s$-channel
and $t$-channel) and diboson ($WW$, $WZ$, and $ZZ$) production.  Each sample is
produced with one of the following event generators:
\pythiaeightver~\cite{Pythia8}, \sherpaver~\cite{Sherpa},
\mcnlover~\cite{MCNLO,MCNLO1,MCNLO2}, \acerver~\cite{AcerMC},
\herwigver~\cite{Herwig6} or
\powhegver~\cite{POWHEG1,POWHEG2,POWHEG3,POWHEGDY}.  The most consistent set of
available samples was chosen.  The \DY{} process is generated at leading order
so that the sample can also be reweighted to describe the \Zprime{} signal.
The combination of \ttbar{} and single-top-quark production are referred to as
{\em top}.  In some cases the generators are interfaced to the following
external software for parton showering, hadronisation and multiple parton
interactions: \pythiaeight{}, \pythiasixver~\cite{Pythia} or \herwig{} (which
is itself interfaced to \jimmyver~\cite{Jimmy} for multiple parton
interactions).  The tau lepton decay is performed by either \pythiaeight{},
\sherpa{} or \tauola{}~\cite{TAUOLA}. For \pythiaeight{}, the {\em
sophisticated tau decay} option is used, which provides fully modelled hadronic
currents with spin correlations for tau-lepton decays~\cite{Pythia8Taus}.  In
all samples other than those generated with \sherpa{}, final-state photon
radiation is performed by \photos~\cite{PHOTOS}.  The \cteq~\cite{CTEQ6L1} and
\ctten~\cite{CT10} parton distribution functions (PDFs) and the \autwo{},
\auettwo{}, \auettwob{}~\cite{ATL-PHYS-PUB-2011-008} and \ctten~\cite{Sherpa}
MC tunes are used.  A summary is given in table~\ref{tab:samples}. 
\end{sloppypar}

The contributions from simulated processes are normalised using theoretical
cross sections.  The \DYonly{} cross section is calculated up to
next-to-next-to-leading order (NNLO) in QCD including next-to-leading order
(NLO) electroweak corrections using \fewzthreever~\cite{FEWZ3} configured with
the \mstwnnlo{} PDF set~\cite{Martin:2009}. This cross section is used to
derive mass-dependent $K$-factors that are used to weight the simulated
$\DYonly{}$ samples.  Cross sections for the other background processes are
calculated without the use of differential $K$-factors to at least NLO in QCD, as
specified in table~\ref{tab:samples}.

\begin{table}[tbp]
\centering
\footnotesize
\begin{tabular}{lllllll}
\toprule
Process          & Generator        & PS+MPI                 & Tau decay    & PDF set & MC tune      & Cross section                   \\
\midrule                                                                    
\DY              & \pythiaeight     & \pythiaeight           & \pythiaeight & \cteq   & \autwo{}     & NNLO~\cite{FEWZ3}               \\
\Wjets           & \sherpa          & \sherpa                & \sherpa      & \ctten  & \ctten       & NNLO~\cite{WZprod,FEWZ2.0}      \\ 
\ttbar{}         & \mcnlo           & \herwig                & \tauola      & \ctten  & \auettwo{}   & $\sim$NNLO~\cite{ttbarxsec1,ttbarxsec2,HATHOR}  \\ 
Single top       \\                                                         
~~($Wt$)         & \mcnlo           & \herwig                & \tauola      & \ctten  & \auettwo{}   & $\sim$NNLO~\cite{tchanxsec}     \\ 
~~($s$-channel)  & \mcnlo           & \herwig                & \tauola      & \ctten  & \auettwo{}   & NNLL~\cite{schanxsec}           \\ 
~~($t$-channel)  & \acer            & \pythiasix             & \tauola      & \cteq   & \auettwob{}  & $\sim$NNLO~\cite{tchanxsec}     \\ 
\Diboson{}       & \herwig          & \herwig                & \tauola      & \cteq   & \auettwo{}   & NLO~\cite{dibosonxsec}          \\
\DYll{}          & \powheg          &\pythiaeight            & \pythiaeight &\ctten   &\autwo        & NNLO~\cite{FEWZ3}               \\ 
\bottomrule
\end{tabular}
\caption{
  Details regarding the MC simulated samples. The following information is provided for each sample: the generator
  of the hard interaction, the parton shower and hadronisation (PS), multiple parton interactions (MPI) and the 
  tau decay; the PDF set; the MC tune and the order in QCD of the cross section calculation. All cross sections 
  are calculated at either NLO, NNLO, approximate NNLO ($\sim$NNLO) or next-to-next-to-leading logarithm (NNLL).
}
\label{tab:samples}
\end{table}

The contributions of the various $\Zprime{}$ signal models are estimated by
reweighting the \DY{} sample using
\tauspinner~\cite{TauSpinner,TauSpinnerVal,TauSpinnerBSM}, which correctly
accounts for spin effects in the tau decays. The algorithm relies on a leading
order approximation in which spin amplitudes are used to calculate the spin
density matrices for hard $2\to2$ Born level processes.  The impact of
interference between \Zprime{} and \DYonly{} is typically small (as discussed
in section~\ref{sec:sig_acc}), so it is not included.  For each signal model,
several mass hypotheses are considered, ranging from 500 to $2500\GeV$ in steps
of $125\GeV$.  

All generated events are propagated through a detailed \geant{}
simulation~\cite{geant4} of the ATLAS detector and subdetector-specific
digitisation algorithms~\cite{AtlasSimulation} and are reconstructed with the
same algorithms as the data.  Pile-up is simulated by overlaying minimum-bias
interactions generated with \pythiapileupver{} (with an MC tune specific to the
LHC~\cite{AMBT2}) on the generated signal and background events.  The resulting
events are reweighted so that the distribution of the number of minimum-bias
interactions per bunch crossing agrees with data. Due to the high momenta of
the tau decay products, however, pile-up has little impact on the analysis.
The effective luminosity of most simulated samples is at least as large as the
integrated luminosity of the data; the statistical uncertainty from the limited
sample size is accounted for in the statistical analysis.

%% file: reco.tex
\section{Physics objects}
\label{sec:reco}

In this section the reconstruction of electrons, muons, hadronic tau decays and
the missing transverse momentum is described. Preliminary selections are
applied to all electrons, muons and tau candidates. Further selection is
applied to some of the objects as part of the {\em event selection} described
in section~\ref{sec:sel}. Corrections are applied to the kinematics and
efficiencies of reconstructed electrons, muons and hadronic tau decays in
simulated samples so that they match the performance measured from the data.

The reconstruction, energy calibration and identification of hadronic tau
decays in ATLAS is described in detail in ref.~\cite{TauPaper}. Candidates for
hadronic tau decays are built from jets reconstructed using the
\mbox{anti-$k_{t}$} algorithm~\cite{anti-kt,anti-kt-fast} with a radius
parameter value of 0.4.  The jets are calibrated to the hadronic energy scale
with correction factors based on simulation and validated using test-beam and
collision data~\cite{ATLASJES}.  Only the {\em visible} tau-decay products (all
products excluding neutrinos), \tauhvis{}, are considered when calculating
kinematic properties. The calculation of the four-momentum uses clusters with
$\Delta R<0.2$ from the initial jet-axis and includes a final tau-specific
calibration derived from simulated samples, which accounts for out-of-cone
energy, energy lost in dead material, underlying-event and pile-up
contributions and the typical composition of hadrons in hadronic tau decays.
The size of the tau-specific calibration is typically a few percent. The calibrated
energy scale in data and simulation have been compared and agree within the
$\sim$1.5\% uncertainty of the measurement.  Candidates are required to have
either one or three associated tracks (prongs) reconstructed in the inner
detector.  The tau charge is reconstructed from the sum of the charges of the
associated tracks and is required to be~$\pm1$. The charge misidentification
probability is found to be negligible.  Hadronic tau decays are identified with
a multivariate algorithm that employs boosted decision trees (BDTs) to
discriminate against quark- and gluon-initiated jets using shower shape and
tracking information.  Working points with a tau identification (ID) efficiency
for 1-prong/3-prong candidates of about 55\%/40\% ({\em medium}) for the
$\muhad$ and $\ehad$ channels and 65\%/45\% ({\em loose}) for the \hadhad{}
channel are chosen, leading to rates of false identification for quark- and
gluon-initiated jets of below a percent.  The tau ID efficiency is independent
of \pt{} and pile-up.  Corrections of a few percent are applied to the efficiency
in simulation. Candidates arising from the misidentification of
electrons are rejected using a separate BDT.  In the \muhad{} channel, a
dedicated selection is applied to suppress candidates arising from the
misidentification of muons.  Tau candidates are required to have \mbox{$\pt{} >
30\GeV$} and to be in the fiducial volume of the inner detector, \mbox{$|\eta|
< 2.47$}.  The transition region between the barrel and endcap EM calorimeters,
with \mbox{$1.37 < |\eta| <1.52$}, is excluded.  In the $\lephad{}$ channels,
candidates that have the highest-\pt{} track in the range \mbox{$|\eta|<0.05$}
are rejected.  This region corresponds to a gap in the TRT, which reduces the
power of electron/pion discrimination. 

Muon candidates are reconstructed by combining an inner-detector track with a
track from the muon spectrometer. The candidates are required to have
\mbox{$\pt{} > 10\GeV$} and \mbox{$|\eta| < 2.5$}. Muon quality criteria are
applied to achieve a precise measurement of the muon momentum and reduce the
misidentification rate~\cite{ATLASMuonPerformance}.  These quality requirements
correspond to a muon reconstruction and identification efficiency greater than
95\%.

Electrons are reconstructed by matching clustered energy deposits in the EM
calorimeter to tracks reconstructed in the inner
detector~\cite{ATLASElectronEnergyCalib}. The tracks are then refitted using
the Gaussian Sum Filter algorithm~\cite{ATLASElectronGSF}, which accounts for
energy loss through bremsstrahlung.  The electron candidates are required to
have \mbox{$\pt{} > 15\GeV$} and to be within the fiducial volume of the inner
detector, \mbox{$|\eta| < 2.47$} (the EM calorimeter transition region is
excluded).  The candidates are required to satisfy quality criteria based on
the expected calorimeter shower shape and amount of radiation in the TRT.
These quality requirements correspond to an electron identification efficiency
of approximately 95\%~\cite{ATLASElectronEfficiency}. 

Electrons and muons are considered isolated if they are away from large
deposits of energy in the calorimeter and tracks with large \pt{} consistent
with originating from the same vertex.  Lepton isolation is defined using the
sum of the transverse energy, \et{}, deposited in calorimeter cells with
$\Delta R < 0.2$ from the lepton, $\et^{0.2}$, and the scalar sum of the \pt{}
of tracks with $\pt > 0.5\GeV$ consistent with the same vertex as the lepton
and with $\Delta R < 0.3$, $\pt^{0.3}$.  Muons are considered isolated if they
have $\pt^{0.3}/\pt < 5\%$.  Isolated electrons must have $\pt^{0.3}/\pt < 5\%$
and $\et^{0.2} < 5\GeV + 0.7\% \times\pt$ and must pass a tighter
identification requirement corresponding to an efficiency of approximately
70\%. 

Geometric overlap of objects with \mbox{$\Delta R < 0.2$} is resolved by
selecting only one of the overlapping objects in the following order of
priority: muons, electrons, tau candidates.  The order is determined by the
ability to identify the objects from their detector signatures. The missing
transverse momentum, with magnitude \met{}, is calculated from the vector sum
of the transverse momenta of all high-\pt{} objects reconstructed in the event,
as well as a term for the remaining activity in the
calorimeter~\cite{AtlasMET2010}.  Clusters associated with electrons, hadronic
tau decays and jets are calibrated separately.  The remaining clusters are
weighted using tracking information to reduce the effect of pile-up on the
\met{} resolution. A single weight is calculated for each event using all
tracks that are not matched to high-\pt{} objects. The tracks are categorised
based on whether or not they are matched to the primary vertex. The weight is
then defined as the ratio of the sum of the \pt{} of tracks originating from
the primary vertex to the sum of the \pt{} of all tracks.

%% file: sel.tex
\section{Event selection}
\label{sec:sel}

\begin{sloppypar}
Selected events in the \hadhad{} channel must contain no electrons with
\mbox{$\pt>15\GeV$} or muons with \mbox{$\pt>10\GeV$} and at least two tau
candidates: one with \mbox{$\pt>150\GeV$} that is matched to the object that
passed the trigger and the other with $\pt>50\GeV$.  This constitutes the {\em
preselection}.  If multiple tau candidates are selected, the two highest-\pt{}
candidates are chosen. This decision is made before applying the BDT tau ID, to
avoid kinematic biases in control regions defined by reversing the ID
requirement.  The tau candidates are then required to have charges of opposite
sign (OS).  Finally, the angle between the tau candidates in the transverse
plane, $\Delta\phi(\tau_1,\tau_2)$, must be greater than 2.7\,radians, as tau
leptons from the decay of heavy neutral resonances are typically produced
back-to-back in the transverse plane. 
\end{sloppypar}

\begin{sloppypar}
Selected events in the $\lephad$ channels must contain exactly one isolated
muon with \mbox{$\pt>30\GeV$} or one isolated electron with
\mbox{$\pt>30\GeV$}; no additional electrons with \mbox{$\pt{}>15\GeV$} or
muons with \mbox{$\pt{}>10\GeV$}; and at least one tau candidate with
\mbox{$\pt{}>30\GeV$}. This constitutes the {\em preselection}.  If multiple
tau candidates are selected, the highest-\pt{} candidate is chosen. As in the
\hadhad{} channel, this choice is made before applying the BDT tau ID. The
angle between the lepton and tau candidate in the transverse plane,
\mbox{$\Delta\phi(\ell,\tau)$}, must be greater than 2.7\,radians, and they
must have opposite charge.  
The {\em transverse mass} is defined as:
\begin{equation*} 
m_{\rm T}(p^{A},p^{B})=\sqrt{2\pt^{A}\pt^{B} \left ( 1-\cos\Delta\phi(p^{A},p^{B}) \right )} \, , 
\end{equation*}
where $p^{A}$ and $p^{B}$ are two reconstructed physics objects with transverse
momenta $\pt^{A}$ and $\pt^{B}$, respectively, which subtend an angle of
$\Delta\phi(p^{A},p^{B})$ in the transverse plane.  The \mbox{\Wjets{}}
background is suppressed by requiring the transverse mass of the lepton-\met{}
system, $\mT(\ell,\met)$, to be less than $50\GeV$.
\end{sloppypar}

The search in all channels is performed by counting events in signal regions
with {\em total transverse mass} above thresholds optimised separately for each
signal mass hypothesis in each channel to give the best expected exclusion
limits. The same thresholds are found to be optimal for all channels.
The total transverse mass, \mTtot, is defined as 
\begin{equation*}
\mTtot(\tau_1,\tau_2,\met) = \sqrt{\mt^2(\tau_1,\tau_2) + \mt^2(\tau_1,\met) + \mt^2(\tau_2,\met)}\, , 
\end{equation*}
where $\tau_1$ and $\tau_2$ denote the reconstructed visible decay products of
the two tau leptons ($e$, $\mu$ or \tauhvis).

%% file: bkg.tex
\section{Background estimation}
\label{sec:bkg}

The dominant background process in the \hadhad{} channel at high mass is \DY{},
which is estimated using simulation. The modelling of the $pp\to\DYonly{}$
process has been shown to be very reliable by using decays to electrons and
muons~\cite{ATLASdilepton,ATLASdrellyann}. Additional uncertainties related to
the modelling of high-\pt{} tau decays are also considered, as described in
section~\ref{sec:sys}.  \Multijet{} production makes a large
contribution at low mass in the \hadhad{} channel and is estimated by weighting
events in data where the subleading tau candidate fails tau ID, with {\em
\ffactors} that parameterise the rate for jets to pass tau ID
(section~\ref{sec:bkg:ffactorhadhad}).  Due to the relatively large size of the
sample that fails tau ID, this procedure provides high statistical precision,
which is particularly crucial in the high-mass tail.  The \ffactors{} are
measured from data in a separate control region. \Diboson{}, \Wjets{},
\ttbar{}, \Zlljets{}, and single-top-quark production make minor
contributions and are estimated using simulation. To improve the modelling of
these background processes, events in the simulation that contain jets
misidentified as hadronic tau decays are weighted by {\em \frates{}} measured
in a \Wjets{} control region in data (section~\ref{sec:bkg:fratehadhad}).

The dominant background contributions in the \lephad{} channels come from
\DY{}, which is estimated using simulation, and from processes in which a jet
is misidentified as a hadronic tau decay. The latter is mainly composed of
\Wjets{} events and is estimated using \ffactors{} to weight events in data
where the tau candidate fails ID, similarly to the procedure in the \hadhad{}
channel (section~\ref{sec:bkg:ffactorlephad}).  \Diboson{}, \ttbar{},
\Zlljets{} and single-top-quark production in which the tau candidate does not
originate from a jet make minor contributions and are estimated using
simulation.  In the following subsections, the data-driven background estimates
are described in more detail.

\subsection{Multijet background in the \hadhad{} channel}
\label{sec:bkg:ffactorhadhad}

To estimate the multijet background in the \hadhad{} channel, two control
regions are used. Events in the first control region are required to pass the
same selection as in the analysis, except for the subleading tau candidate,
which is required to fail the BDT tau ID.  The multijet contribution in the
signal region is estimated by weighting these events with a tau ID \ffactor{}.
The \ffactor{}, \ftauid{}, is defined as the ratio of the number of tau
candidates that pass the BDT tau ID, \Npasstau{}, to the number that fail,
\Nfailtau{}.  The \ffactors{} are calculated from a second control region that
is highly pure in multijet events, the {\em multijet control region} (described below), and
they depend on the \pt{} and track multiplicity, \numTrack{}, of the
subleading tau candidate: 
\begin{equation*}
\ftauid(\pt,\numTrack) \equiv \left.\frac{\Npasstau{}(\pt, \numTrack)}{\Nfailtau(\pt, \numTrack)}\right|_\mathrm{multijet} \, .
\end{equation*}
The \ffactors{} have no significant dependence on $\eta$.
The number of multijet events in a bin of \pt{}, \numTrack{} and any additional variable
that is uncorrelated to the BDT tau ID, $x$, is given by: 
\begin{equation*}
N_\mathrm{multijet}(\pt,\numTrack, x)  =  \ftauid(\pt,\numTrack) \times \Nfailtaudata(\pt, \numTrack, x) \, .
\end{equation*}

The multijet control region is designed to be as similar to the signal region
as possible, while avoiding contamination from hadronic tau decays. This is
achieved by loosening the tau ID requirements. Specifically, the selection for
this control region is the same as for the signal region except with the
following alterations.  The BDT tau ID is not applied to either tau candidate.
Instead of using the single-tau trigger, events are selected using single-jet
triggers with thresholds ranging from 45 to $360\GeV$, with all but the highest
threshold trigger being prescaled.  The \pt{} of the subleading tau candidate
must be at least 40\% of the \pt{} of the leading tau candidate
(\pt{}-balance $>0.4$) to avoid bias at low \pt{} due to the disproportionate
fraction of events coming from the unprescaled $360\GeV$ jet trigger.  The
opposite-sign requirement on the charges of the two tau candidates is removed
to increase the sample size.

\begin{figure}[htbp]
\centering
\includegraphics[width=0.49\textwidth]{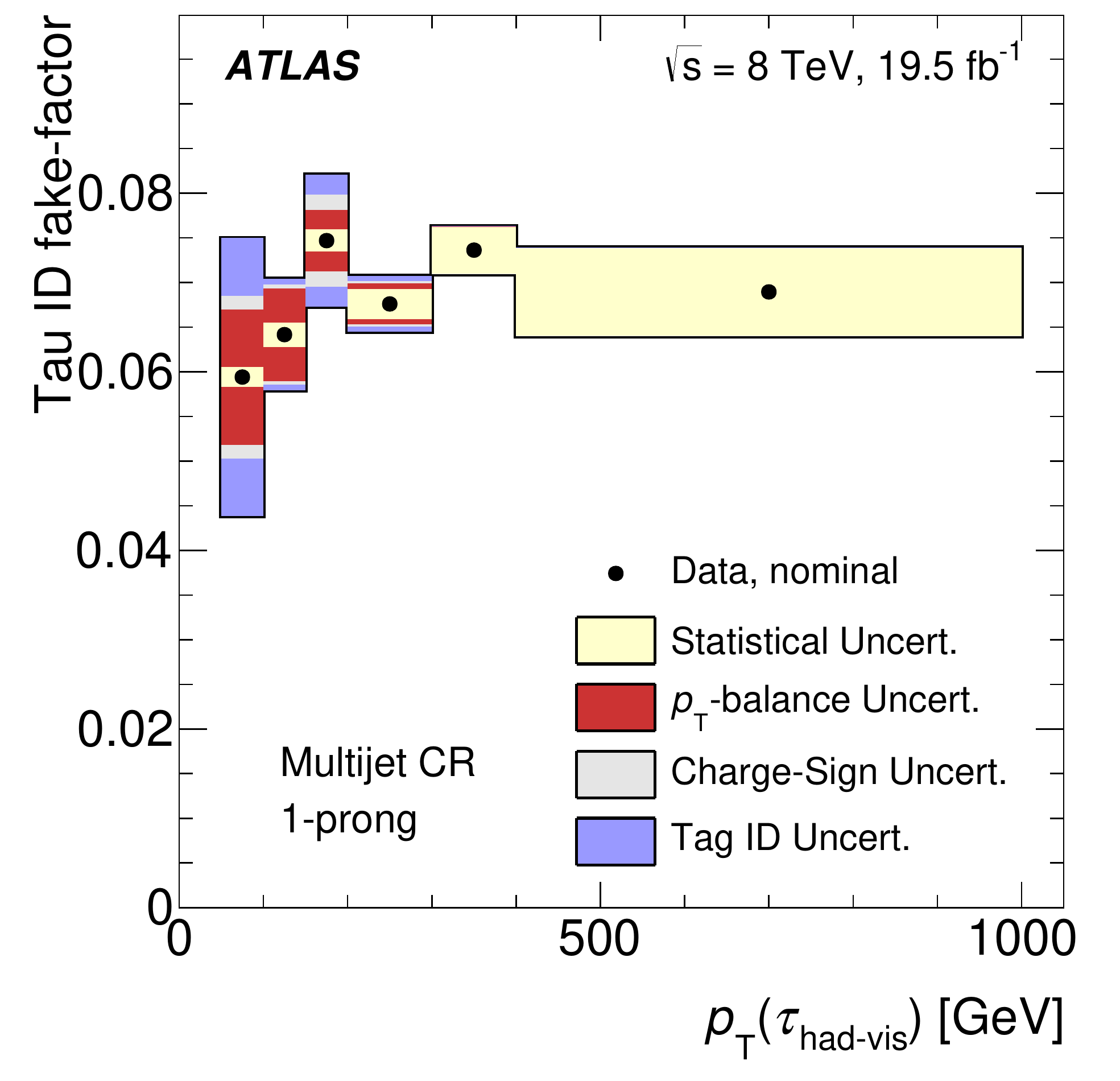}
\includegraphics[width=0.49\textwidth]{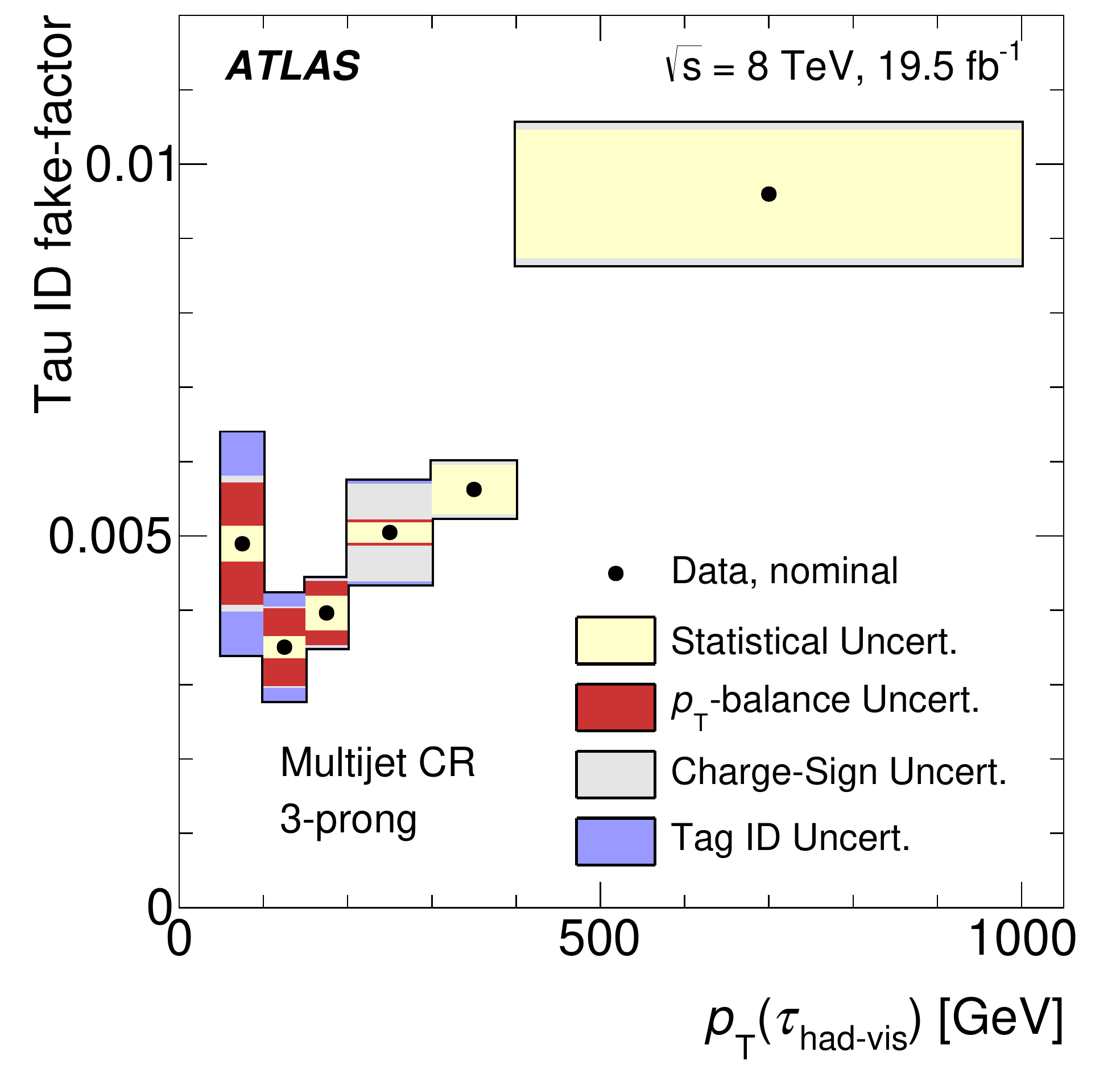}
\caption{Tau ID \ffactors{} for (left) 1-prong and (right) 3-prong 
tau candidates, measured in the multijet control region of the \hadhad{} channel. 
The statistical and systematic uncertainties are shown, successively added in quadrature. }
\label{fig:hadhad:fake-factors}
\end{figure}

Figures~\ref{fig:hadhad:fake-factors}(left) and
\ref{fig:hadhad:fake-factors}(right) show the \ffactors{} for 1-prong and
3-prong candidates, respectively.  Use of these \ffactors{} relies on the
assumption that they are insensitive to the alteration of the selection between
the signal region and multijet control region.  Systematic uncertainties on the
\ffactors{} are derived by altering the selection on the \pt{}-balance, the
charge product, and the identification of the leading-\pt{} tau candidate.
These variations modify the fractional contribution of quark- and
gluon-initiated jets in the sample, leading to large variations in the
\ffactors{} at low \pt{} where the composition is mixed and little variation to
the \ffactors{} at high \pt{} where the sample is quark dominated.

\subsection{Jet background in the \lephad{} channel}
\label{sec:bkg:ffactorlephad}

\begin{figure}[htbp]
\centering
  \includegraphics[width=0.49\textwidth]{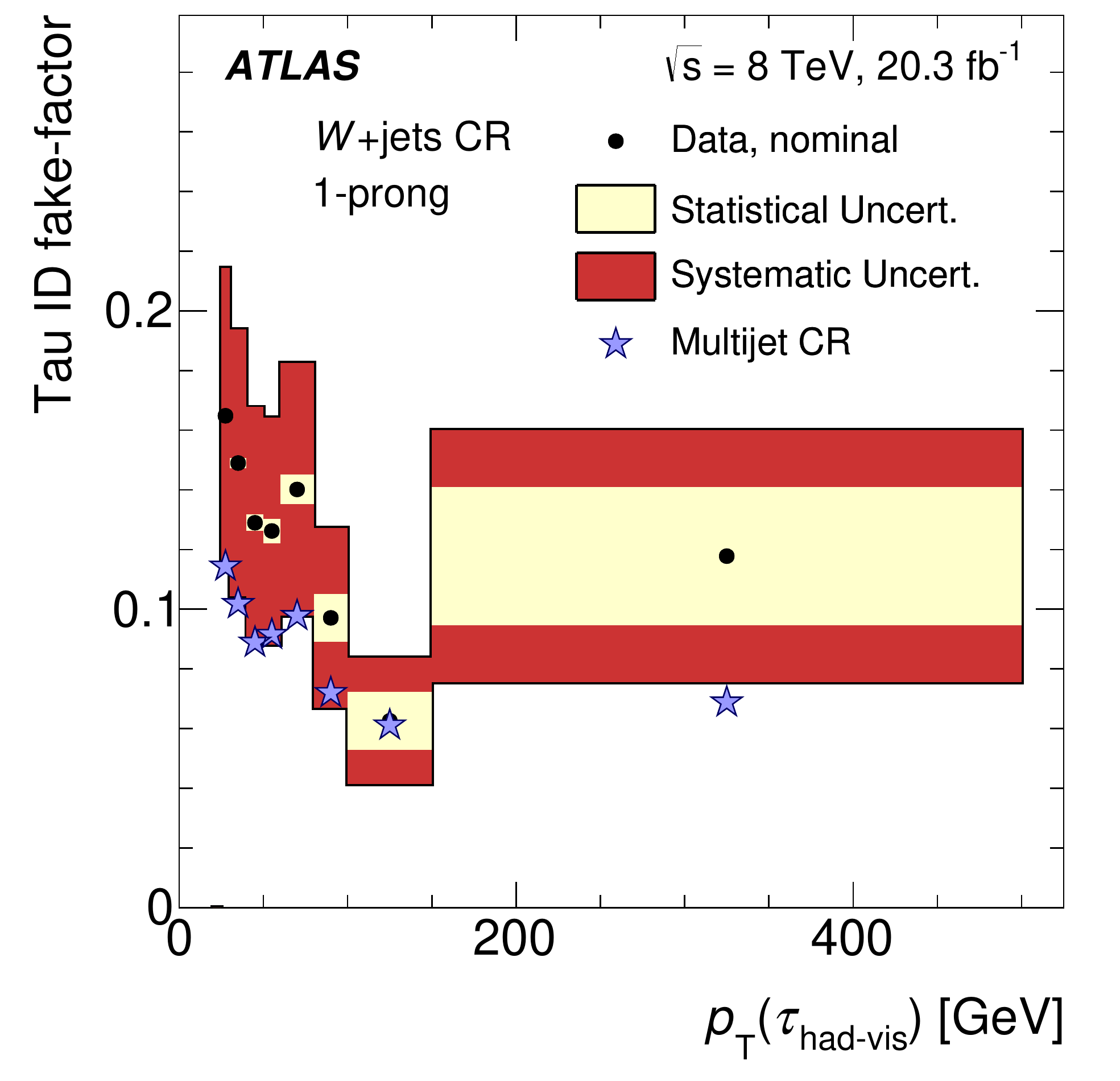}
  \includegraphics[width=0.49\textwidth]{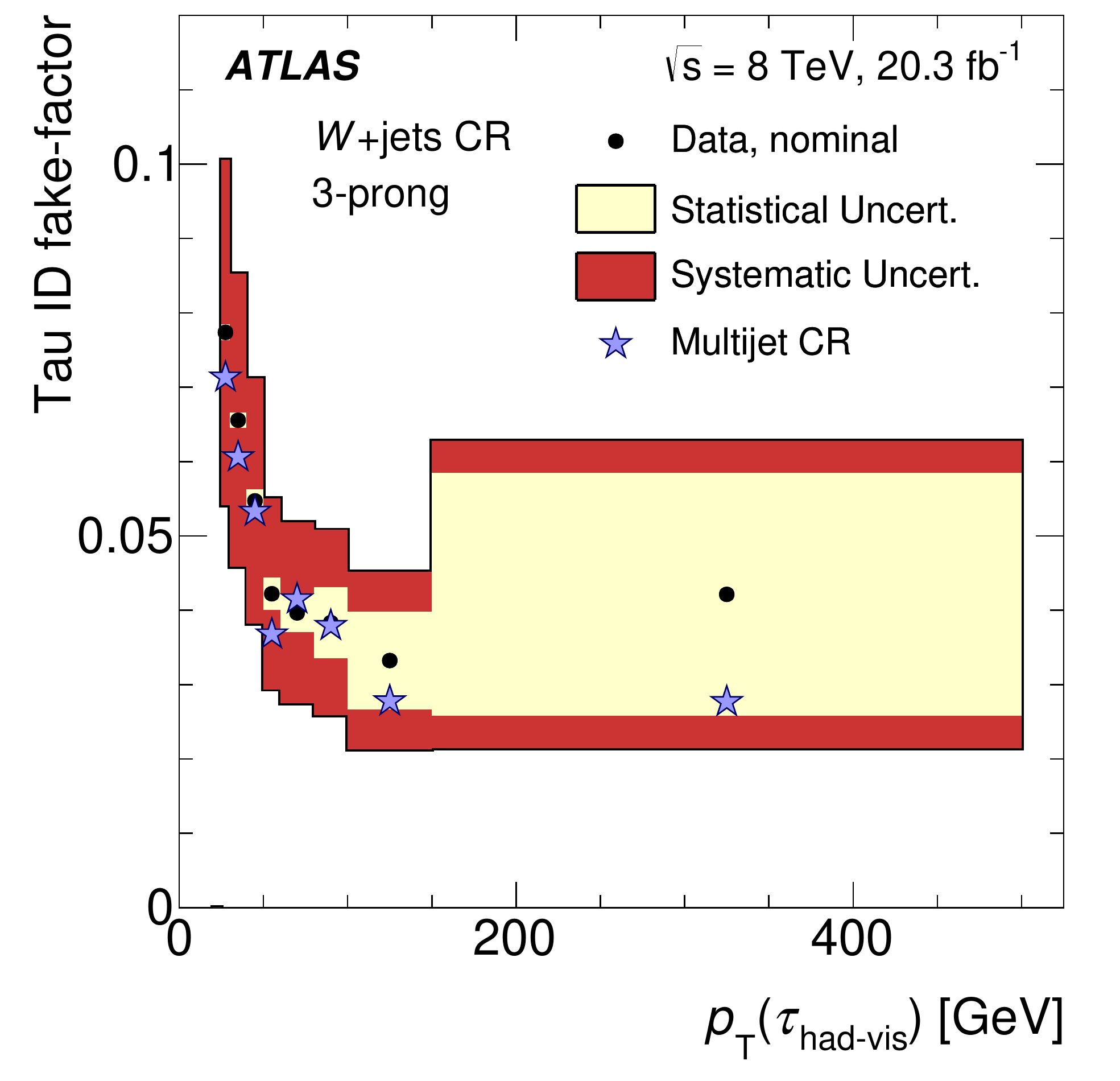}
\caption{
    Tau ID \ffactors{} for (left) 1-prong and (right) 3-prong tau candidates,
    measured in the \Wjets{} control region of the \muhad{} channel, integrated
    across all $|\eta|$ regions.  The statistical and systematic uncertainties are
    shown, successively added in quadrature. The \ffactors{} measured in the alternative
    \multijet{} control region are overlaid.
}
\label{fig:lephad:fake-factors}
\end{figure}

The background contributions originating from quark- and gluon-initiated jets
that are misidentified as hadronic tau decays in the \lephad{} channels are
modelled using a \ffactor{} method, similar to that used in the \hadhad{}
channel. In contrast to the \hadhad{} channel, the background is dominated by
\Wjets{} production, with a minor contribution from \multijet{} production.  To
reduce the sensitivity to the differing \ffactors{} in \Wjets{} and \multijet{}
events (due to a different quark/gluon fraction), events failing a {\em very
loose} level of BDT tau identification (corresponding to efficiencies of
$~$98\% and $~$90\% for 1-prong and 3-prong hadronic tau decays, respectively)
are rejected.  This significantly suppresses the gluon contribution, which
typically consists of wider jets with higher hadron multiplicity which are more
readily rejected by the tau ID.  In the \lephad{} channels, there is also a
non-negligible contribution to the first control region (fail-ID control
region) from background processes containing hadronic tau decays, which is
subtracted using simulation.  The \ffactors{} are measured in a high-purity
\Wjets{} control region and they depend on \pt{}, \eta{} and \numTrack{} of the
tau candidate.  The \Wjets{} control region uses the same selection as the
signal region but with the medium BDT tau ID replaced by very loose BDT tau ID
and the $\mT$ requirement replaced by $70\GeV\leq\mT\leq200\GeV$.  A second
control region enriched in \multijet{} events is defined, which has a higher
fraction of gluon-initiated jets and represent an extreme variation in the jet
composition.  This control region uses the same selection as the \Wjets{}
control region but the lepton is required to fail isolation, the \mT{}
requirement is replaced by $\mT<30\GeV$ and $\metscalar<30\GeV$ is required.  A
30\% systematic uncertainty is derived from the difference in the \ffactors{}
in the multijet and \Wjets{} control regions.
Figures~\ref{fig:lephad:fake-factors}(left) and
\ref{fig:lephad:fake-factors}(right) show the \ffactors{} measured in each of
the two control regions in the \muhad{} channel, integrated across all $|\eta|$
regions, for 1-prong and 3-prong candidates, respectively.  The \ffactors{} in
the \ehad{} channel are similar.  

Finally, in the \lephad{} channel, two additional steps are taken to ensure
\met{} is modelled well by the \ffactor{} estimate. Firstly, the standard
\met{} reconstruction treats the selected tau candidate in the fail-ID control
region as a jet rather than a hadronic tau decay. Therefore, the \met{} is
recalculated in the fail-ID control region using the tau hypothesis for the
selected tau candidate.  Following this, a slight bias in the shape of the
\met{} distribution is corrected for by reweighting in bins of the \met{}
projected along the direction of the tau candidate. 
An additional 20\% uncertainty is applied to the estimate of the jet background
event yield obtained after the full event selection, derived from the
difference in the estimate between applying and not applying the
\met{} reweighting. 

\subsection{Jet background other than multijet in the \hadhad{} channel}
\label{sec:bkg:fratehadhad}

\newcommand{\Rtauidone}{\mbox{\ensuremath{R_\text{tau-ID}^{\rm lead}}}}
\newcommand{\Rtauidtwo}{\mbox{\ensuremath{R_\text{tau-ID}^\text{sub-lead}}}}
\newcommand{\Rtauidi}{\mbox{\ensuremath{R_\text{tau-ID}^{i}}}}
\newcommand{\Ntottau}{\mbox{\ensuremath{ N^{\rm total}}}}
\newcommand{\Npasstautrig}{\mbox{\ensuremath{ N^{\text{pass tau-ID + trigger}}}}}
\newcommand{\typeone}{\mbox{\ensuremath{\mathcal{T}^{\rm lead}}}} 
\newcommand{\typetwo}{\mbox{\ensuremath{\mathcal{T}^\text{sub-lead}}}}

In the \hadhad{} channel, backgrounds originating from quark- and
gluon-initiated jets that are misidentified as hadronic tau decays in processes
other than \multijet{} production are estimated using simulation (predominantly
\Wjets{}).  Rather than applying the tau ID to the simulated jets, they are
weighted by \frate{}s. This not only ensures the correct \frate{}, but enhances
the statistical precision of the estimate, as events failing the tau ID are not
discarded.  The \frate{} for the sub-leading tau candidate, \Rtauidtwo{}, is
defined as the ratio of the number of tau candidates that pass tau ID,
\Npasstau{}, to the total number of tau candidates, \Ntottau{}.  The \frate{}
for the leading tau candidate, \Rtauidone{}, is defined as the ratio of the
number of tau candidates that pass tau ID and the single-tau trigger
requirement, \Npasstautrig{}, to \Ntottau{}.  The \frate{}s are calculated from
a second control region that is high in \Wjets{} purity (described below), and
they depend on \pt{} and \numTrack{} of the tau candidate:  
\begin{align*}
\Rtauidone(\pt,\numTrack) &\equiv \left.\frac{\Npasstautrig{}(\pt, \numTrack)}
  {\Ntottau(\pt, \numTrack)}\right|_{\Wjets} \, , \\
\Rtauidtwo(\pt,\numTrack) &\equiv \left.\frac{\Npasstau{}(\pt, \numTrack)}
  {\Ntottau(\pt, \numTrack)}\right|_{\Wjets} \, .
\end{align*}
All simulated events are assigned a weight: 
\begin{equation*}
w_{\rm MC} = \prod_{i\in \{\text{lead, sub-lead}\}} 
  \left( 1 - \delta^{i}\left[1 - \Rtauidi(\pt^{i},\numTrack^{i})\right] \right )
\end{equation*}
where $\delta^{i}$ is 1 if the tau candidate originates from a jet and 0 otherwise.
The tau ID and trigger selection criteria for simulated events are modified as
follows: the BDT tau ID criteria for the sub-leading tau candidate is removed
if the candidate originates from a jet, the BDT tau ID criteria for the leading
tau candidate and the trigger requirement are removed if the leading tau
candidate originates from a jet.

\begin{figure}[htbp]
  \centering
  \includegraphics[width=0.49\textwidth]{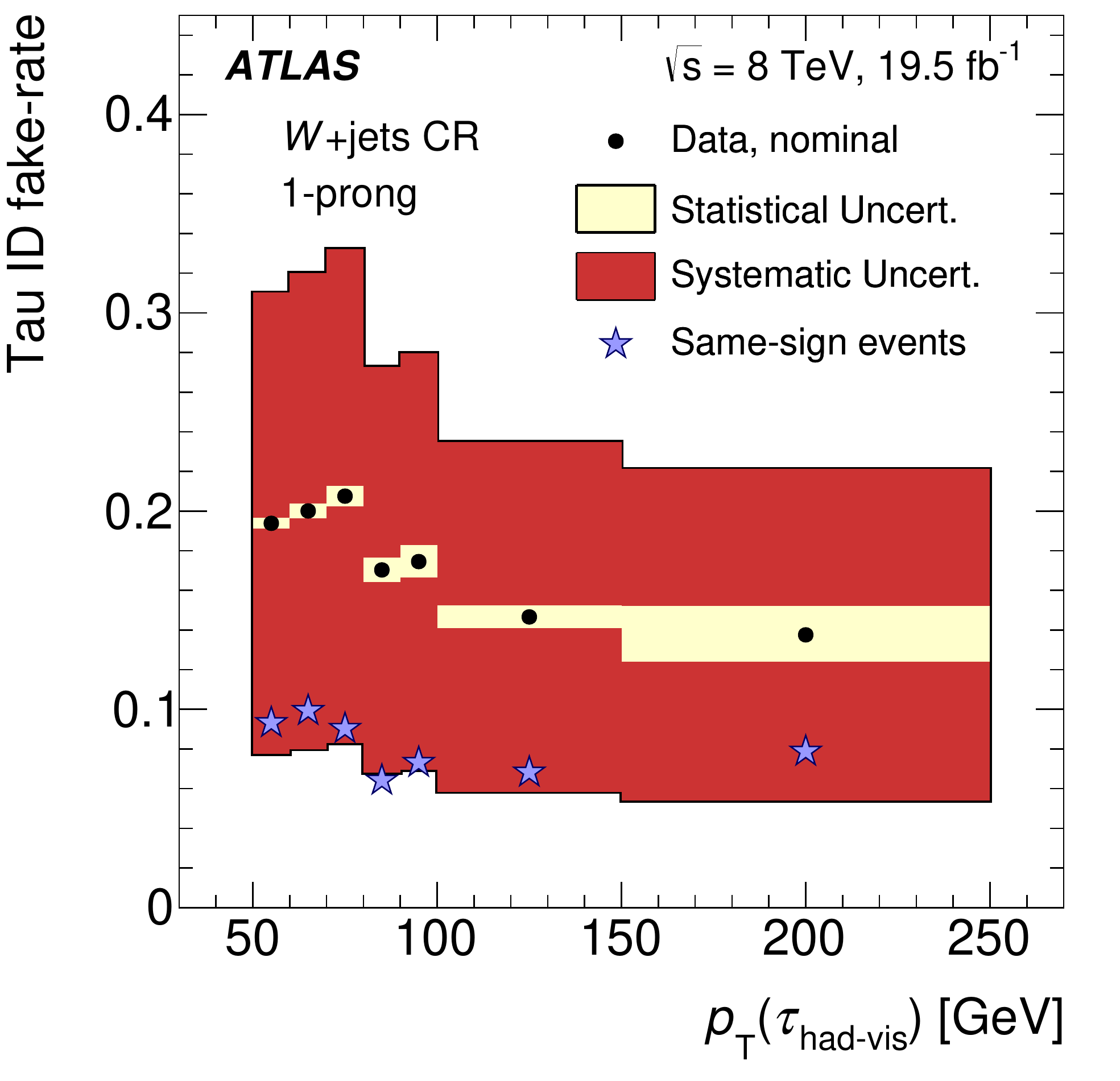}
  \includegraphics[width=0.49\textwidth]{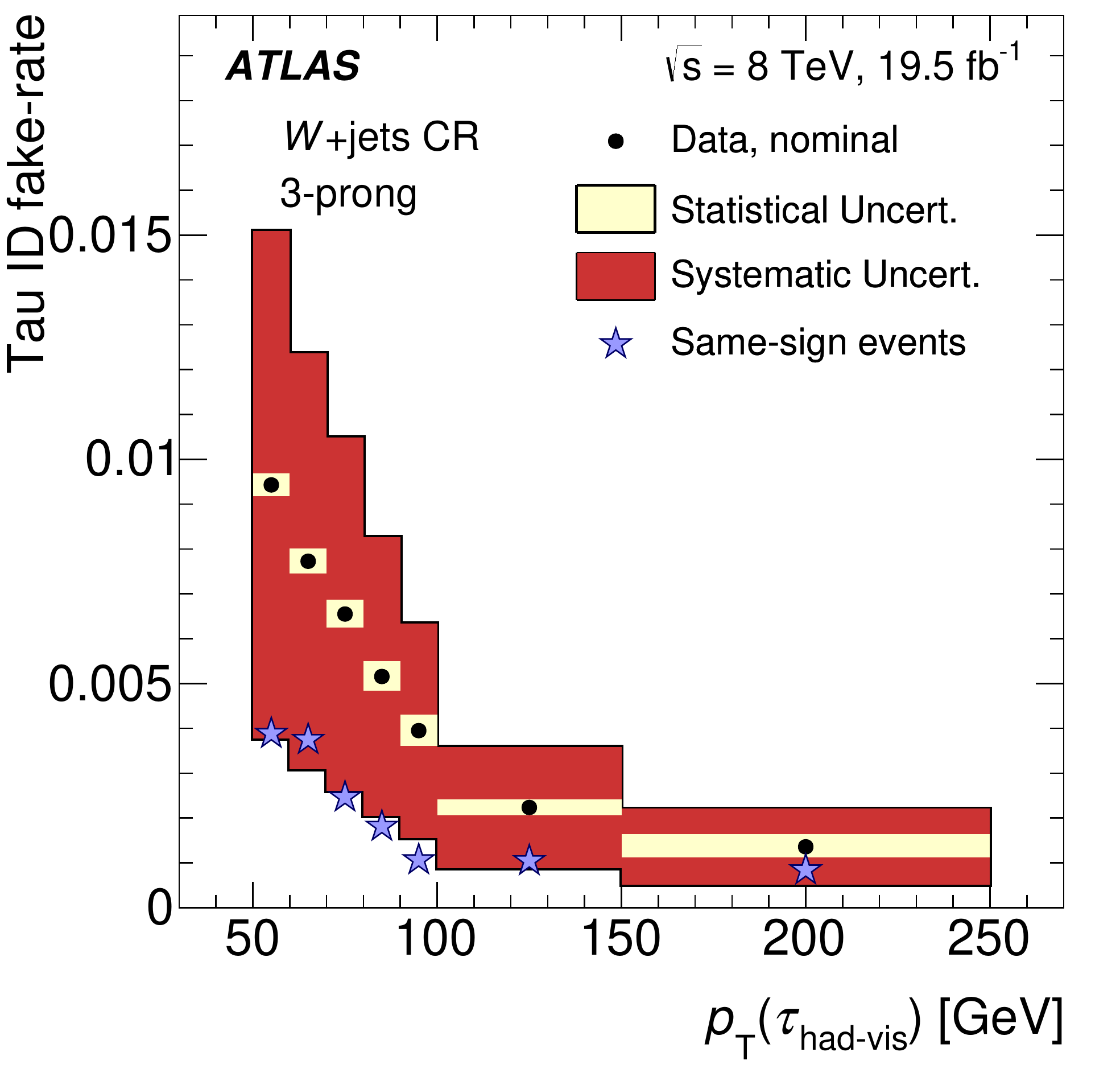}
  \caption{Tau identification \frate{} measured in \Wmunujets{} data events
           for the BDT loose identification working point for (left) 1-prong and (right) 3-prong
           tau candidates. The \frate{} is parameterised in the charge product of
           the muon and fake tau candidate. Opposite-sign events are depicted by black 
           circles and same-sign events by blue stars. The systematic uncertainty 
           covers differences due to jet composition and is added to the statistical 
           uncertainty in quadrature.}
  \label{fig:fakerate:chargeproduct}
\end{figure}

Events in the \Wjets{} control region are selected by a single-muon trigger
with a \pt{} threshold of $36\GeV$.  The events are required to contain one
isolated muon that: has $\pt>40\GeV$, has $\et^{0.2}/\pt<6\%$ and  is matched
to the object that passed the trigger. There must be no additional muons or
electrons and at least one tau candidate with opposite charge to the muon. The
remaining contamination from multijet events is suppressed by requiring
\mbox{$\cos\Delta\phi(\mu,\met)+\cos\Delta\phi(\tauhvis,\met)<-0.15$}, which
disfavours back-to-back topologies where the \met{} vector points either in the
direction of the muon or the tau candidate.  The leading-\pt{} tau candidate is
used to measure the \frate{}.  Figures~\ref{fig:fakerate:chargeproduct}(left)
and \ref{fig:fakerate:chargeproduct}(right) show \Rtauidtwo{} for 1-prong and
3-prong tau candidates, respectively.  The \frates{} \Rtauidone{} (including
the trigger requirement in the numerator) have a similar behaviour but are a
factor of two to four lower.  The requirement of opposite charge between the
muon and the tau candidate enhances the contribution of the leading-order
$qg\to W\bar{q}^{\prime}$ process in which the tau candidate originates from a
quark-initiated jet.  To evaluate the systematic uncertainty from applying
these \frate{}s to simulated samples with different jet origin, the \frates{}
are also calculated for events where the tau candidate has the same charge sign
as the muon. These events have a higher fraction of gluon-initiated jets and
represent an extreme variation in the jet compostion, resulting in lower
\frate{}s as shown in Fig.~\ref{fig:fakerate:chargeproduct}(left) and
Fig.~\ref{fig:fakerate:chargeproduct}(right). A 60\% uncertainty is assigned to
cover the range of the measured \frates{} for events with opposite- or
same-sign tau candidates.  The uncertainty is omitted for \Wjets{} events as
they are expected to have the same jet composition as events in the control
region. The statistical uncertainty from the limited size of the \Wjets{}
control region is also considered.

%% file: sys.tex
\section{Systematic uncertainties}
\label{sec:sys}

Systematic effects on the contributions of signal and background processes
estimated from simulation are discussed in this section.  These include
theoretical uncertainties on the cross sections of simulated processes and
experimental uncertainties on the trigger, reconstruction and identification
efficiencies; on the energy and momentum scales and resolutions; and on the
measurement of the integrated luminosity.  Uncertainties on the background
contributions estimated from data are discussed in their respective sections.

The overall uncertainty on the \Zprime{} signal and the \hbox{$\DYonly\to
ee/\mu\mu/\tau\tau$} background due to choice of the PDFs, $\alpha_{\rm S}$,
and the renormalisation and factorisation scales is estimated to be 14\% for a
ditau mass of $1750\GeV$, dominated by the PDF
uncertainty~\cite{ATLASdilepton}.  The uncertainty is evaluated using 90\%~CL
\mstwnnlo{} PDF error sets and also takes into account potential differences
between the following PDFs at the same $\alpha_{\rm S}$: MSTW2008NNLO,
CT10NNLO, NNPDF2.3~\cite{Ball:2012cx}, ABM11~\cite{Alekhin:2012ig} and
HERAPDF1.5~\cite{Aaron:2009aa}.  Additionally, for \DY{}, a mass-dependent
systematic uncertainty of up to 4\% is attributed to electroweak
corrections~\cite{ATLASdilepton}.  This uncertainty is not considered for the
signal as it is strongly model dependent.  An uncertainty of 5\% is estimated
for \diboson{} production, derived from scale, PDF and $\alpha_{\rm S}$
variations.  A 6\% uncertainty on the \Wjets{} normalisation is derived from
comparisons to data in the \Wjets{} control region used to measure jet-to-tau
fake-rates in the \hadhad{} channel.  For \ttbar{} and single-top-quark
production, the uncertainties from variations in the renormalisation and
factorisation scales are in the range of
3--6\%~\cite{tchanxsec,PhysRevD78034003,Beneke}, while those related to the
proton PDFs amount to
8\%~\cite{CT10,Martin:2009,Botje:2011sn,2011arXiv1101.0536A,Ball:2011mu}.

The uncertainty on the integrated luminosity is 2.8\%.  It is derived from a
preliminary calibration of the luminosity scale derived from beam-separation
scans performed in November 2012, following the same methodology as that
detailed in ref.~\cite{lumi2011}.  Comparisons of the efficiency of the
hadronic tau trigger measured in data and in simulation are used to derive an
uncertainty of 10\% on the trigger efficiency.  Differences between data and
simulation in the reconstruction and identification efficiency and the energy
scale of hadronic tau decays are taken into account.  The associated
uncertainties for muons and electrons are negligible for this analysis.  

The systematic uncertainty on the identification efficiency of hadronic tau
decays is estimated at low \pt{} from data samples enriched in \Ztautau{}
events, yielding an uncertainty of 2--7\% depending on the number of tracks and
$|\eta|$ of the tau candidate. At high \pt{}, there are no abundant sources of
real hadronic tau decays from which an efficiency measurement could be made.
Rather, the tau identification is studied in high-\pt{} dijet events as a
function of the jet \pt{}, which indicates that there is no degradation in the
modelling of the detector response as a function of the \pt{} of tau
candidates.  Based on the limited precision of these studies, an additional
uncertainty of \mbox{$14\%\cdot\pt$/TeV} for 1-prong tau candidates and
\mbox{$8\%\cdot\pt$/TeV} for 3-prong tau candidates is added in quadrature to
the low-\pt{} uncertainty for candidates with \mbox{$\pt>100\GeV$}.  The
reconstruction efficiency for 3-prong tau candidates decreases at high \pt{}
due to track merging. An uncertainty of \mbox{$50\%\cdot\pt$/TeV} above
\mbox{$\pt=150\GeV$} is assigned for 3-prong candidates, derived from data/MC
comparisons of tracking performance within jets.  The energy scale uncertainty
for hadronic tau decays and jets is evaluated based on the single-hadron
response in the calorimeters~\cite{ATLASJES,TauPaper}. In addition, the tau
energy scale is validated in data samples enriched in \Ztautau{} events.  The
systematic uncertainty related to the tau energy scale is a function of $\eta$,
\pt{} and the number of prongs, and is generally near 3\%. Energy scale and
resolution uncertainties for all objects are propagated to the \met{}
calculation.  The uncertainty on the \met{} due to clusters that do not belong
to any reconstructed object has a minor effect. 

Table~\ref{tab:finalsys} summarises the systematic uncertainties across all
channels for the \mbox{$1750\GeV$} \ZprimeSSM{} mass point. In the \hadhad{} channel
the dominant uncertainties on both the signal and background come from the tau
efficiency and energy scale, while in the \lephad{} channels the statistical
uncertainty on the background coming from the fake-factor estimate also makes a
major contribution.  The uncertainties are the same for background and similar
for the signal for all higher signal mass points, since the same \mTtot{}
thresholds are used.  The uncertainties for the lower mass points are typically
very similar, except for the tau ID efficiency, the 3-prong tau reconstruction
efficiency, the \DYonly{} cross section and the statistical uncertainties,
which are all a few percent lower, and the uncertainty on the tau energy scale
for the signal, which can be up to 11\% at low mass since the \mTtot{}
requirement is much tighter relative to the \Zprime{} mass.  The small
data-driven uncertainty contribution to the signal in the \hadhad{} channel
comes from jets that are misidentified as hadronic tau decays.

\begin{table}[htbp]
    \renewcommand{\tabcolsep}{1mm}
    \small
    \centering
    \begin{tabular}{lrrrrrrrr}
        \toprule
Uncertainty [\%]                    & \hspace{1mm}     & \multicolumn{3}{c}{Signal} &\hspace{3mm}                    & \multicolumn{3}{c}{Background}   \\
                                    &      & {\hadhad} & {\muhad} & {\ehad} &  & {\hadhad}   & {\muhad} & {\ehad} \\ 
        \midrule
        Statistical uncertainty     &      &  2.4 &   4   &  4   &  &  6   & 21   & 21   \\
        Efficiency                  &      & 16   &   8   &  8   &  & 12   &  5   &  4   \\
        Energy scale and resolution &      &  2.9 &   5   &  5   &  &  10  & 11   &  9   \\
        Theory cross section        &      & {--} &  --   & --   &  &  6   &  6   &  6   \\
        Luminosity                  &      &  2.8 &   2.8 &  2.8 &  &  2.5 &  2.2 &  1.9 \\
        Data-driven methods         &      &  0.2 &  {--} & {--} &  &  2.7 &  8   &  12  \\
\midrule
        Total                       &      &  17  &  11   &   10 &  &  18  &  27  &  28  \\
        \bottomrule
    \end{tabular}
\caption{
    Uncertainties on the estimated \ZprimeSSM{} contribution ($m_{\ZprimeSSM}=1750\GeV$) and 
    the corresponding total background contribution in percent for each channel.  
    A dash denotes that the uncertainty is not applicable.  The statistical
    uncertainty corresponds to the uncertainty due to the limited size of the samples 
    produced via simulation or selected in control regions. The total consists
    of all uncertainties added in quadrature. 
}
\label{tab:finalsys}
\end{table}

%% file: models.tex
\section{\Zprime{} signal models}
\label{sec:models}

In this section, the impact on the signal acceptance times efficiency from
altering the \Zprime{} couplings and from including interference between
\Zprime{} and \DYonly{} is discussed.  The acceptance times efficiency 
for a given \Zprime{} model is defined as: 
\begin{equation*}
\acceff = \frac{N_{\rm S}}{\mathcal{L}_{\rm int} \cdot \xsecbr }
\end{equation*}
where $N_{\rm S}$ is the expected number of \Zprime{} events passing the full
analysis selection, \xsecbr{} is the \Zprime{} production cross section times
\tptm{} branching fraction and $\mathcal{L}_{\rm int}$ is the integrated
luminosity.
The impact on \acceff{} is presented as a fraction of the SSM value,
\acceffssm{}.  The corresponding impact on the acceptance alone, \acc{}, is
also evaluated by replacing $N_{\rm S}$ with the expected number of \Zprime{} 
events after applying the kinematic selection directly to the generated
particles before simulation. A \Zprime{} model that couples preferentially to
third-generation fermions is also discussed.

\subsection{\Zprime{} signal acceptance} 
\label{sec:sig_acc}

\begin{figure}[htbp]
\centering
\includegraphics[width=0.49\textwidth]{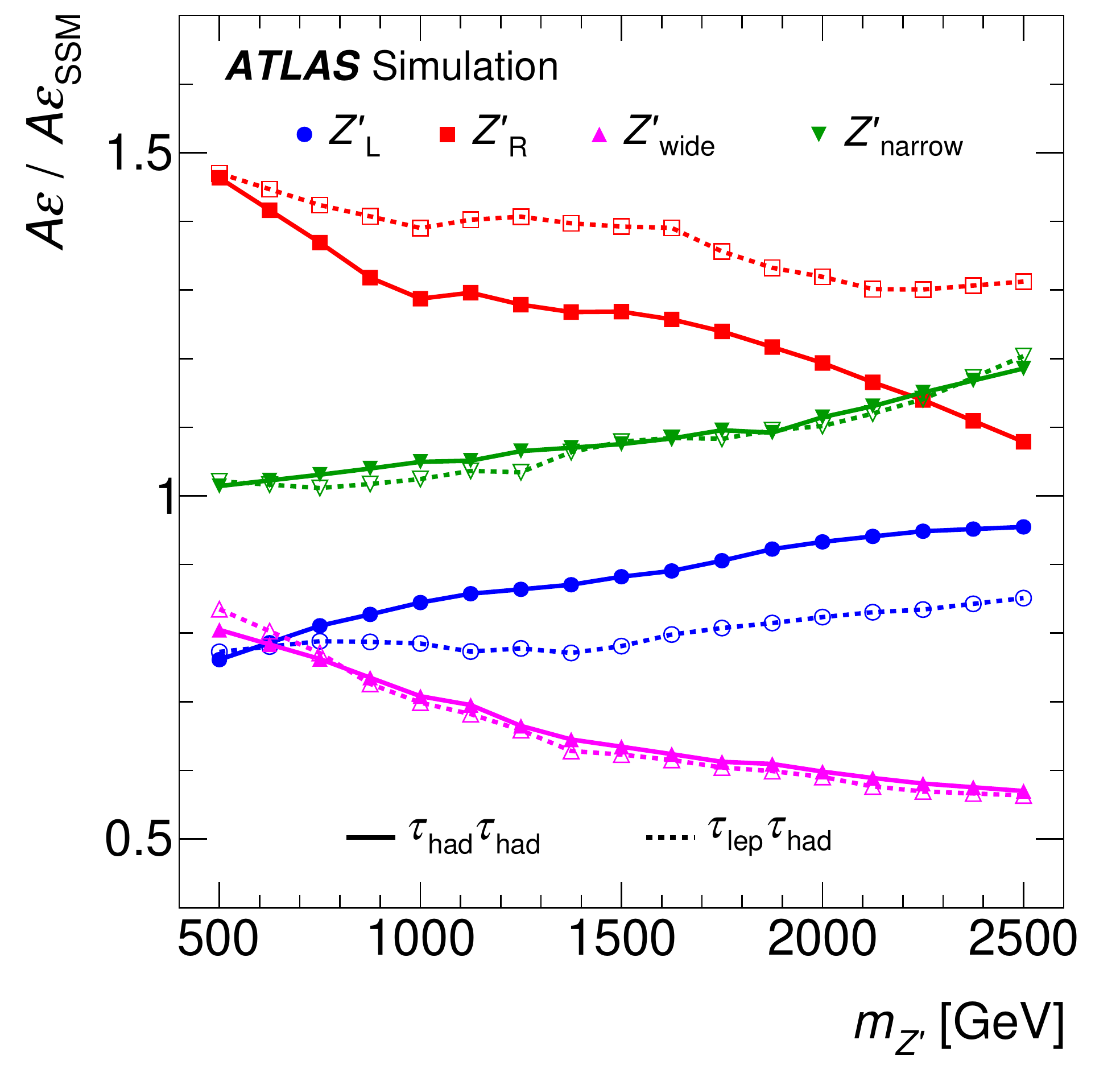}
\caption{
  Signal acceptance times efficiency for \ZprimeL{}, \ZprimeR{},
  \ZprimeNarrow{} and \ZprimeWide{} divided by the acceptance times efficiency
  for \ZprimeSSM{} as a function of $m_{\Zprime}$, separately for the \hadhad{}
  (solid lines with filled markers) and \lephad{} (dashed lines with empty
  markers) channels. The statistical uncertainty is typically below 5\% but can
  increase to 14\% at low mass.
} 
\label{fig:acc}
\end{figure}

Changing the fermionic couplings of the \Zprime{} from their SSM values can
alter the signal acceptance of the analysis. Such changes are primarily due to
alterations in either the tau polarisation or the total \Zprime{} decay width.
Alteration of the tau polarisation changes the tau decay kinematics.  Most
importantly it affects the visible momentum fraction, which enters the analysis
through the \pt{} thresholds of the reconstructed visible tau decay products
and via the threshold on \mTtot{}. The most extreme impact on the acceptance is
seen for models that couple only to left-handed or right-handed tau leptons:
\ZprimeL{} and \ZprimeR{}, respectively. Alteration of the quark couplings can
impact the acceptance if it alters the tau polarisation. However, the maximum
impact is much smaller than when altering the couplings to tau leptons. As the
kinematic limit (due to the collision energy) for high-mass \Zprime{}
production is approached, the signal exhibits an increased fraction of low-mass
off-shell production.  The fraction of off-shell events increases rapidly as a
function of the decay width.  Figure~\ref{fig:acc} shows \acceff{} for the
\ZprimeL{} and \ZprimeR{} models, and two models with artificially altered
decay widths: \ZprimeNarrow{} ($\Gamma/m_{\Zprime}=1\%$) and \ZprimeWide{}
($\Gamma/m_{\Zprime}=20\%$), each divided by \acceff{} for \ZprimeSSM{}
($\Gamma/m_{\Zprime}\approx3\%$).  Interference between \Zprime{} and \DYonly{}
is not included. The statistical uncertainty is typically below 5\% but can be
up to 14\% at low mass. A smoothing is applied to reduce fluctuations.  For
\ZprimeL{} and \ZprimeR{}, the largest impact is observed at low mass, where
the \pt{} and \mTtot{} thresholds are much more stringent on the signal.  In
this case, alteration of the tau couplings can lead to changes of up to $+50\%$
and $-25\%$.  The impact on the \hadhad{} and \lephad{} channels are different
due to the different effect of polarisation on leptonic and hadronic tau
decays.  For \ZprimeNarrow{} and \ZprimeWide{}, the impact is most prominent at
high mass where changes of up to $+20\%$ and $-45\%$ are observed. At low mass,
\acceff{} only changes for widths above 10\%. The impact is the same for all
channels.  For all \Zprime{} models, the change in \acc{} is very similar to
that in \acceff{}, indicating that the efficiency is insensitive to changes in
the \Zprime{} couplings.

The impact of interference between \Zprime{} and \DYonly{} is typically small.
For the SSM, it leads to a reduction in the expected \Zprime{} contribution of
up to 10\% for $m_{\Zprime}\le2\TeV$, and up to $35\%$ for the highest mass
hypotheses. For \ZprimeL{}, \ZprimeR{} and \ZprimeNarrow{} the impact is
negligible. For \ZprimeWide{} the impact can be substantial and is highly
dependent on the choice of the fermionic couplings. An exhaustive treatment is
outside the scope of this article. Reinterpretations of the SSM results for
models with large widths should specifically calculate the impact from
interference. 

\subsection{Non-universal $G(221)$ model}   
\label{sec:sig_sfm}

The non-universal $G(221)$ model~\cite{SFM,Topflavor,G221} (also 
known by other names such as {\em Topflavor}) is an
extension of the SM, containing additional heavy gauge bosons, 
\ZprimeSFM{} and \WprimeSFM{}, that may couple preferentially to
third-generation fermions.  The model is motivated by the idea that the large
mass of the top-quark may suggest that the third fermion generation has a dynamical
behaviour different from the first two generations. Accordingly, the SM weak
$SU(2)$ gauge group is split into two parts: one coupling to {\em light} fermions
(the first two generations), $SU(2)_{\rm l}$ and one coupling to {\em heavy} fermions
(the third generation), $SU(2)_{\rm h}$. The extended gauge group breaks to the SM
$SU(2)_{\rm l+h}$ at a high energy scale, $u$, and then eventually to $U(1)_{\rm EM}$ at the
usual electroweak scale, $v=246\GeV$: 
\begin{equation*}
SU(2)_{\rm l} \times SU(2)_{\rm h} \times U(1)_Y \overset{u}{\rightarrow} SU(2)_{\rm l+h} \times U(1)_Y \overset{v}{\rightarrow} U(1)_{\rm EM}\,.
\end{equation*}
The mixing between $SU(2)_{\rm l}$ and $SU(2)_{\rm h}$ is described by the parameter
$\sin^{2}\phi$.  The \ZprimeSFM{} and \WprimeSFM{} bosons are degenerate in
mass; the mass is defined at tree level by $\sin^2\phi$ and $u$. Large mixing between
$\tau$ and $\mu$ leptons has been considered as an additional feature of the
model, but is ignored here as it would lead to stronger limits via the
dielectron and dimuon searches.  The \ZprimeSFM{} couples almost exclusively to
left-handed fermions, and while the coupling strength differs for light
and heavy fermions, it is largely insensitive to the electric charge or
weak isospin, leading to almost universal couplings for all light and heavy fermions.

\begin{figure}[htbp]
\centering
\includegraphics[width=\textwidth]{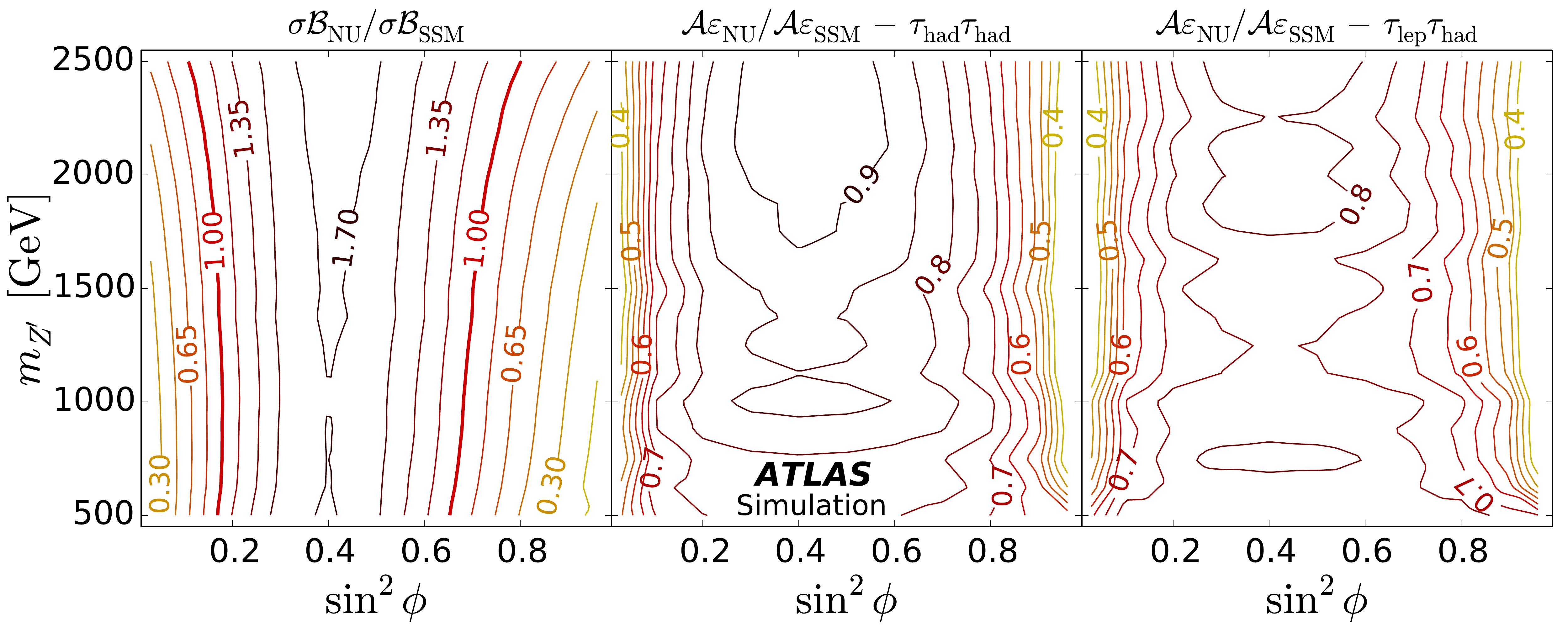}
\caption{
  Signal production cross section times $\tau^+\tau^-$ branching fraction for \ZprimeSFM{}, 
  \xsecbrsfm{}, divided by \xsecbrssm{} (left) and acceptance times efficiency for
  \ZprimeSFM{}, \acceffsfm{}, divided by \acceffssm{} for the (middle) \hadhad{} 
  and (right) \lephad{} channels, as a function of $\sin^2\phi$ and $m_{\Zprime}$.
} 
\label{fig:sfm_xsec_acc_ratios}
\end{figure}

Figure~\ref{fig:sfm_xsec_acc_ratios}(left) shows the \ZprimeSFM{} cross section
times \tptm{} branching fraction, \xsecbrsfm{}, divided by \xsecbrssm{}.  For
much of the parameter space \xsecbrsfm{} is larger than \xsecbrssm{}, peaking
at moderate values of $\sin^2\phi$.  For extreme values of $\sin^2\phi$ either
the cross section is suppressed by weakened couplings to light quarks
($\sin^2\phi\sim0$) or the branching fraction is suppressed by weakened
couplings to tau leptons ($\sin^2\phi\sim1$).
Figures~\ref{fig:sfm_xsec_acc_ratios}(middle) and
\ref{fig:sfm_xsec_acc_ratios}(right) show the \ZprimeSFM{} acceptance times
efficiency, \acceffsfm{}, divided by \acceffssm{}, for the \hadhad{} and
\lephad{} channels, respectively. In general \acceffsfm{} is lower than
\acceffssm{}.  At low mass this is mainly due to the left-handed couplings,
which result in softer visible tau decays.  Near $\sin^2\phi\sim0$ and
$\sin^2\phi\sim1$, the acceptance loss comes mainly from the significantly
increased decay width, which causes a large fraction of the signal to be
produced off shell.  

%% file: results.tex
\section{Results and discussion}
\label{sec:results}

\begin{figure*}[htbp]
    \centering
    \includegraphics[width=0.49\textwidth]{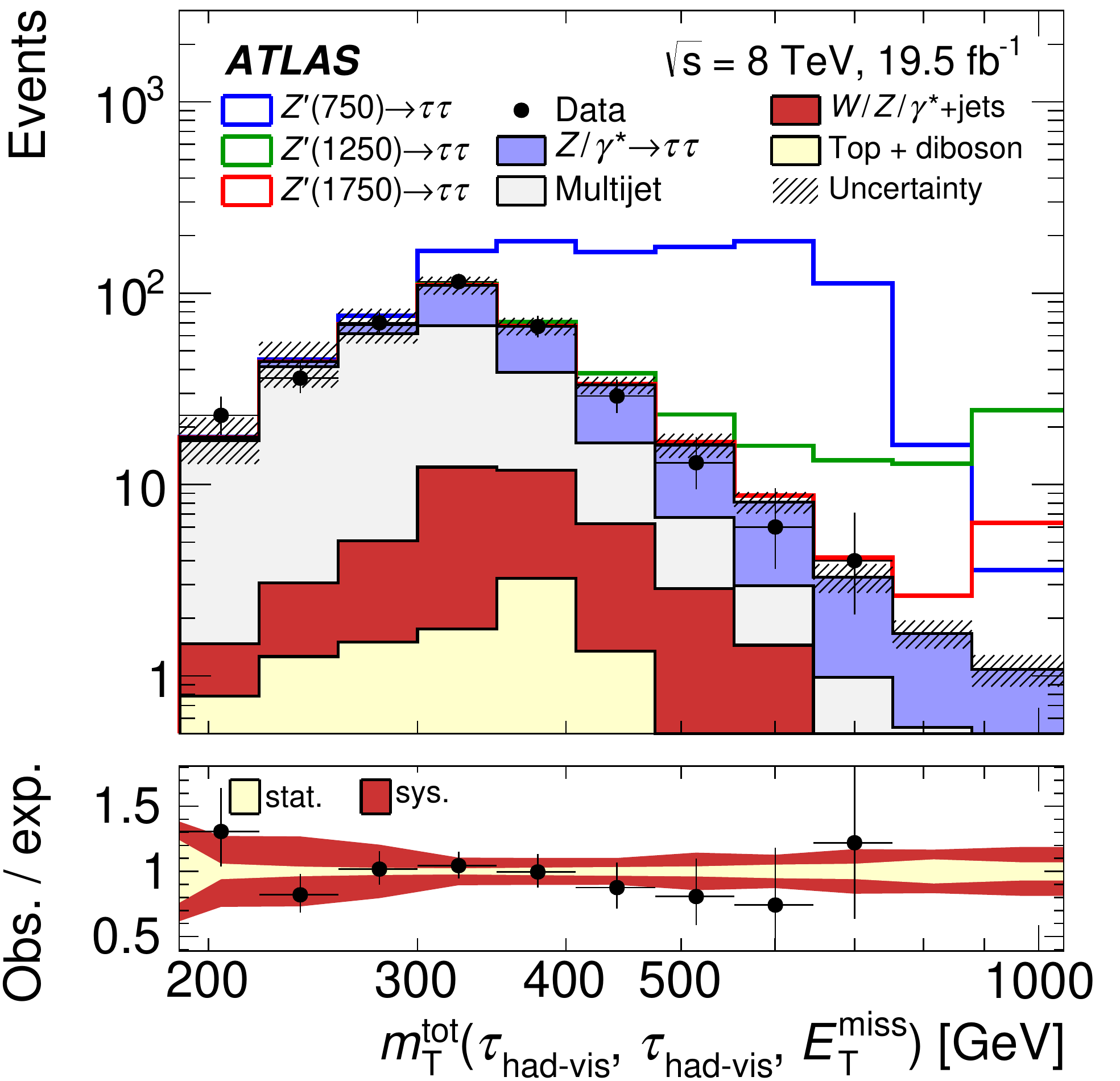}
    \includegraphics[width=0.49\textwidth]{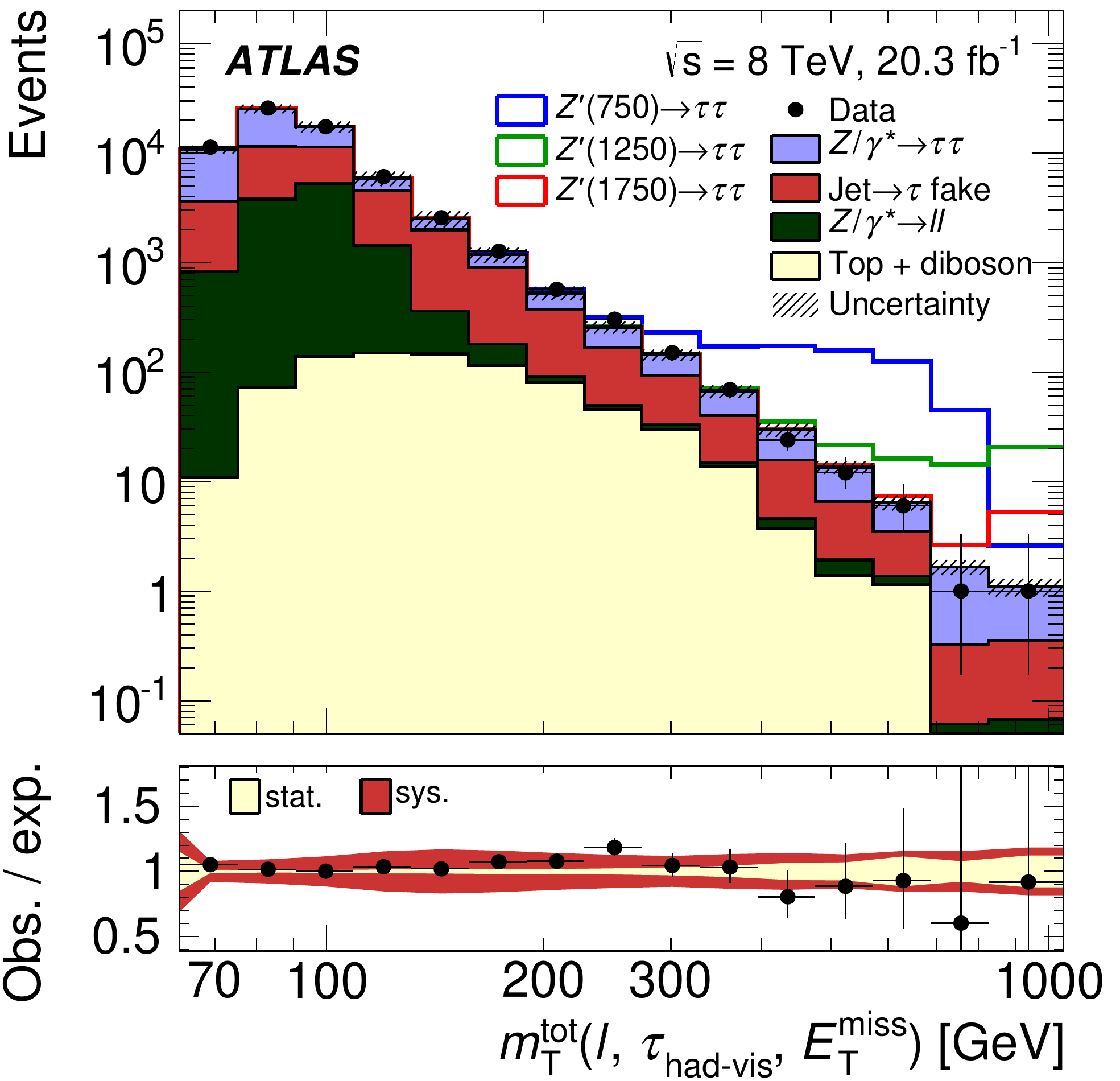}
    \caption{
        The \mTtot{} distribution after event selection in the (left) \hadhad{} 
        and (right) \lephad{} channels. The estimated contributions from
        SM processes are stacked and appear in the same order as in the legend.
        The expected contributions from three \ZprimeSSM{} signals with masses of 
        750, 1250 and $1750\GeV$ are shown, stacked on the total SM expectation.
        The events observed in data are overlaid.  The hatched area indicates the uncertainty 
        on the total estimated background. The bins have a constant width of (left) 0.153
        and (right) 0.184 in $\log(\mTtot)$.
        The last bin includes overflow events.
        The inset shows the ratio of the observed events over the total expected SM
        contribution. The statistical uncertainty from the observed events and the
        expected SM contribution are shown on the points and by the yellow band,
        respectively. The red band depicts the total systematic and statistical uncertainties
        on the SM contribution added in quadrature.
    }
    \label{fig:finalmass}
\end{figure*}

\begin{figure}[htbp]
    \centering
    \includegraphics[width=0.49\textwidth]{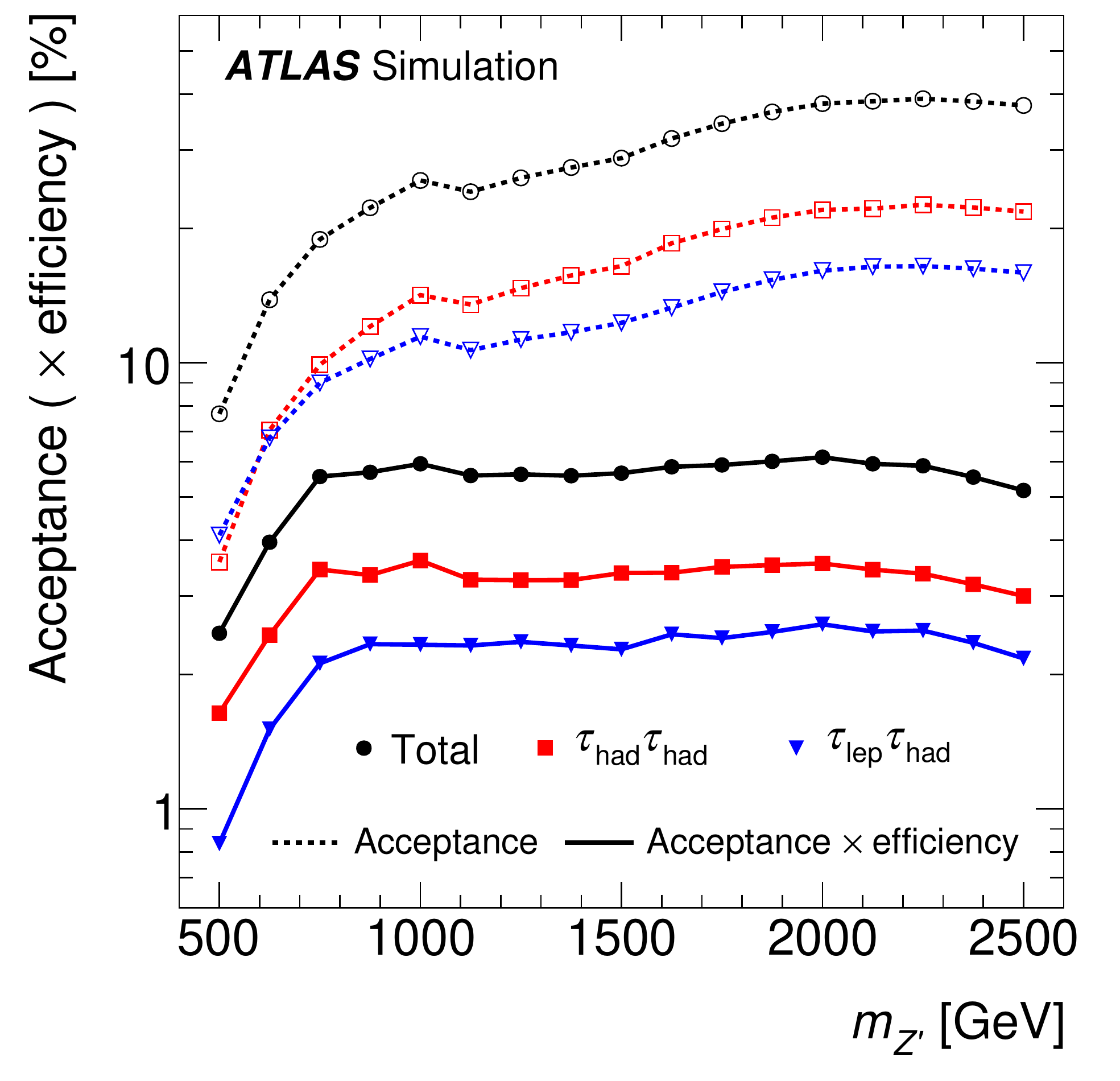}
    \caption{
        Acceptance (dashed lines with empty markers) and acceptance times 
        efficiency (solid lines with filled markers) for \ZprimeSSM{} 
        as a function of the \ZprimeSSM{} mass. Contributions from the 
        individual channels and the full analysis are given.  
    }
    \label{fig:acceptance}
\end{figure}

A summary of the expected number of events remaining after successively
applying each selection requirement, up to the \mTtot{} threshold, for the
signal and dominant background processes is given in table~\ref{tab:cutflow}.
Figures~\ref{fig:finalmass}(left) and \ref{fig:finalmass}(right) show the \MT{}
distribution after event selection in the \hadhad{} and \lephad{} channels,
respectively.  The numbers of observed and expected events (including their
total uncertainties) after applying the \mTtot{} thresholds in all channels are
summarised in table~\ref{tab:eventsummary}. In all cases, the number of
observed events is consistent with the expected Standard Model background.
Therefore, upper limits are set on the production of a high-mass resonance
decaying to \tptm{} pairs.  The acceptance and acceptance times efficiency for
\ZprimeSSM{} is shown in figure~\ref{fig:acceptance}.

\begin{table*}[htbp]
    \centering
    \small
    \sisetup{table-align-uncertainty=true,
             table-number-alignment=center,
             table-figures-decimal=0,
             table-figures-integer=5,
            }
    \begin{tabular}{
      l
      S[]
      S[]
      S[]
      S[]
      S[table-figures-decimal=3,table-figures-integer=2]
    }
    \toprule
\multicolumn{6}{c}{\hadhad{} channel} \\
                                 & {\DY{}}       & {\Multijet{}} & {\WZjets{}} & {Top + \diboson} & {\ZprimeSSM{}} \\ 
\midrule
Preselection                     &  276\pm18    &  611\pm5     &  64\pm1 &  24\pm2   &  10.1\pm0.2    \\
OS                               &  270\pm18    &  316\pm4     &  53\pm1 &  21\pm2   &   9.5\pm0.2    \\
$\Delta\phi(\tau_1,\tau_2)>2.7$  &  117\pm 2    &  209\pm3     &  35\pm1 &  11\pm2   &   9.2\pm0.2    \\

\toprule
\multicolumn{6}{c}{\lephad{} channel} \\
                                 & {\DY{}}     & {Jet $\to\tau$ fake} & {\DYll} & {Top + \diboson} & {\ZprimeSSM{}} \\
\midrule
Preselection                     & 46800\pm300 &    154670\pm130 &  17340\pm250 & 12330\pm70 &  14.3\pm0.2 \\
OS                               & 46300\pm300 &    111270\pm120 &  16180\pm240 & 11830\pm70 &  13.9\pm0.2 \\
$\Delta\phi(\ell,\tau)>2.7$      & 32200\pm300 &     47650\pm80  &  12490\pm210 &  3530\pm40 &  13.5\pm0.2 \\    
$m_{\rm T}<50\GeV$               & 29490\pm230 &     22660\pm60  &  11240\pm210 &   808\pm16 &   8.5\pm0.2 \\   
\bottomrule
\end{tabular}
\caption{
  Number of expected signal ($m_{\ZprimeSSM}=1750\GeV$) and background events in
  the \hadhad{} and \lephad{} channels after successively applying each selection criterion.
  The statistical uncertainty in the least significant digit(s) is shown in parentheses.  
}
\label{tab:cutflow}
\end{table*}

\begin{table*}[htbp]
    \centering
    \sisetup{table-number-alignment=right,table-figures-integer=2,table-figures-decimal=2}
    \begin{tabular}{cc|
      S[table-figures-uncertainty=1,table-figures-integer=4]
      S[table-figures-uncertainty=1]
      S[table-number-alignment=center,table-figures-integer=2,table-figures-decimal=0]|
      S[table-figures-uncertainty=1,table-figures-integer=3,table-figures-decimal=3]
      S[table-figures-uncertainty=1]
      S[table-number-alignment=center,table-figures-integer=2,table-figures-decimal=0]
      }
    \toprule
    $m_{\Zprime}$ & \mTtot{} & \multicolumn{3}{c|}{\hadhad{}}   & \multicolumn{3}{c}{\lephad{}} \\
    \multicolumn{2}{c|}{$[$GeV$]$} & {$N_{\rm S}$} & {$N_{\rm B}$} & $N$ & {$N_{\rm S}$} & {$N_{\rm B}$} & $N$ \\ 
    \midrule
       500&       400&     1030\pm170      &     70  \pm8        &        56&      570\pm90       &     49  \pm6        &       42\\
       625&       450&      650\pm100      &     40  \pm5        &        30&      420\pm50       &     29  \pm4        &       23\\
       750&       500&      410\pm60       &     24.0\pm3.0      &        18&      270\pm29       &     18.2\pm2.3      &       15\\
       875&       550&      206\pm30       &     14.6\pm2.0      &        11&      152\pm14       &     11.2\pm1.6      &       10\\
      1000&       600&      119\pm17       &     9.4 \pm1.3      &         4&     82  \pm8        &     6.7 \pm1.1      &        6\\
      1125&       700&       60\pm9        &     4.0 \pm0.6      &         0&     45  \pm5        &     2.5 \pm0.4      &        2\\
      1250&       750&       35\pm6        &     2.8 \pm0.5      &         0&     27.0\pm2.9      &     1.78\pm0.32     &        1\\
      1375&       800&     20.8\pm3.4      &     1.93\pm0.32     &         0&     15.7\pm1.6      &     1.24\pm0.23     &        1\\
      1500&       850&     13.4\pm2.2      &     1.32\pm0.24     &         0&     9.6 \pm1.0      &    0.96 \pm0.20     &        1\\
      1625&       850&     8.4 \pm1.4      &     1.32\pm0.24     &         0&     6.5 \pm0.7      &    0.96 \pm0.20     &        1\\
      1750&       850&     5.4 \pm0.9      &     1.32\pm0.24     &         0&     4.0 \pm0.4      &    0.96 \pm0.20     &        1\\
      1875&       850&     3.6 \pm0.6      &     1.32\pm0.24     &         0&     2.70\pm0.27     &    0.96 \pm0.20     &        1\\
      2000&       850&     2.4 \pm0.4      &     1.32\pm0.24     &         0&     1.85\pm0.18     &    0.96 \pm0.20     &        1\\
      2125&       850&     1.54\pm0.28     &     1.32\pm0.24     &         0&     1.19\pm0.12     &    0.96 \pm0.20     &        1\\
      2250&       850&     1.02\pm0.19     &     1.32\pm0.24     &         0&    0.81 \pm0.08     &    0.96 \pm0.20     &        1\\
      2375&       850&     0.66\pm0.12     &     1.32\pm0.24     &         0&    0.52 \pm0.05     &    0.96 \pm0.20     &        1\\
      2500&       850&    0.43 \pm0.08     &     1.32\pm0.24     &         0&    0.330\pm0.034    &    0.96 \pm0.20     &        1\\
    \bottomrule
    \end{tabular}
\caption{
    Number of expected \ZprimeSSM{} signal ($N_{\rm S}$), background ($N_{\rm B}$) and observed ($N$)
    events in the \hadhad{} and \lephad{} channels. The signal mass ($m_{\Zprime}$) 
    and corresponding \mTtot{} thresholds are given in units of GeV.
    The total uncertainty (statistical and systematic added in quadrature) in the least significant digit(s) is shown in parentheses.  
}
\label{tab:eventsummary}
\end{table*}

The statistical combination of the channels employs a likelihood function
constructed as the product of Poisson-distributed random numbers describing the
total number of events observed in each channel. The probability in each
channel is evaluated for the observed number of data events given the
signal-plus-background expectation. Systematic uncertainties on the expected
number of events are incorporated into the likelihood via nuisance parameters
constrained by Gaussian distributions. Correlations between signal and
background and across channels are taken into account.  A signal-strength
parameter multiplies the expected signal in each channel, for which a positive
uniform prior probability distribution is assumed. Theoretical uncertainties on
the signal cross section are not included in the calculation of the
experimental limit as they are model dependent.

\begin{figure}[htbp]
    \centering
    \includegraphics[width=0.49\textwidth]{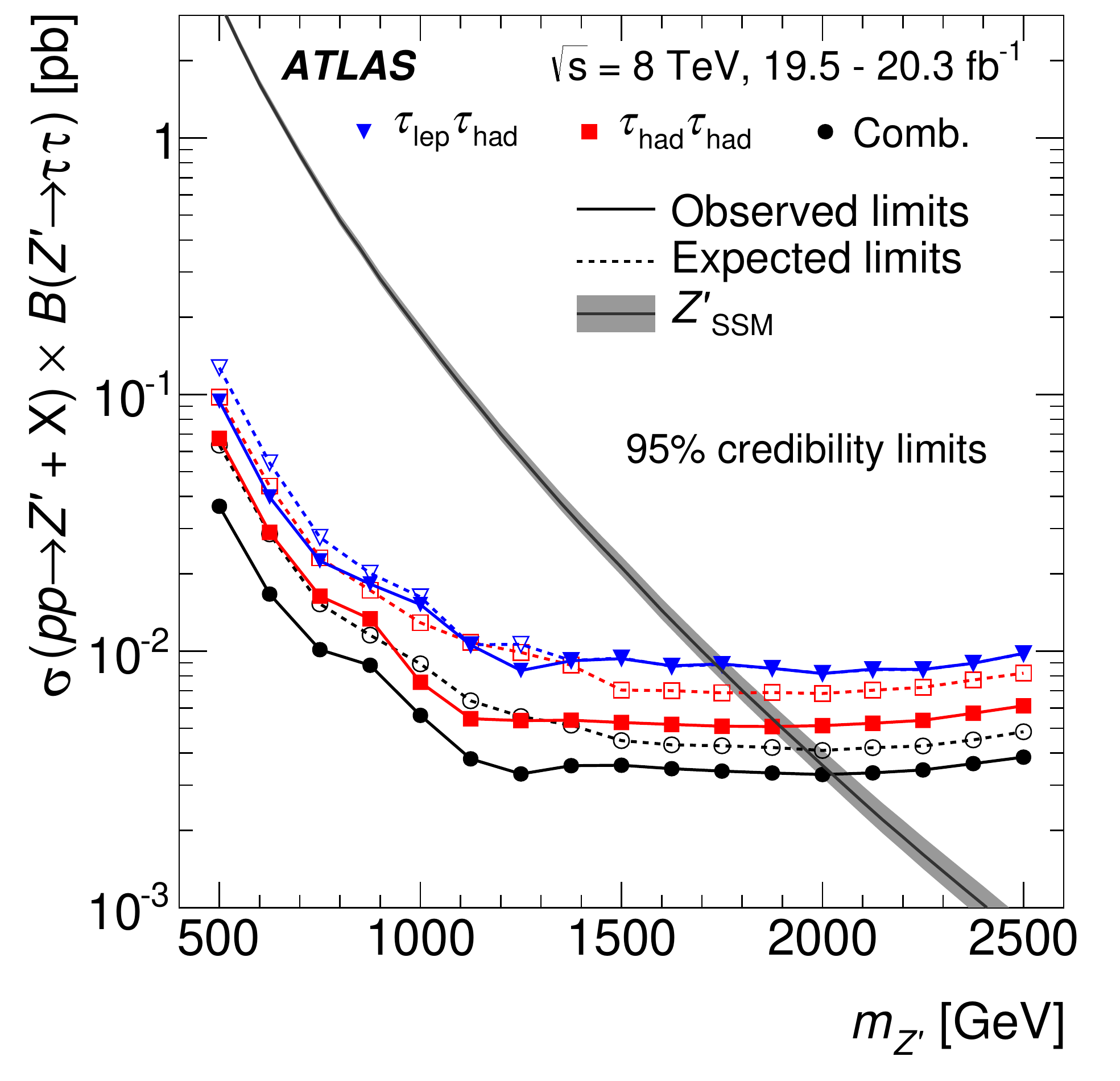}
    \includegraphics[width=0.49\textwidth]{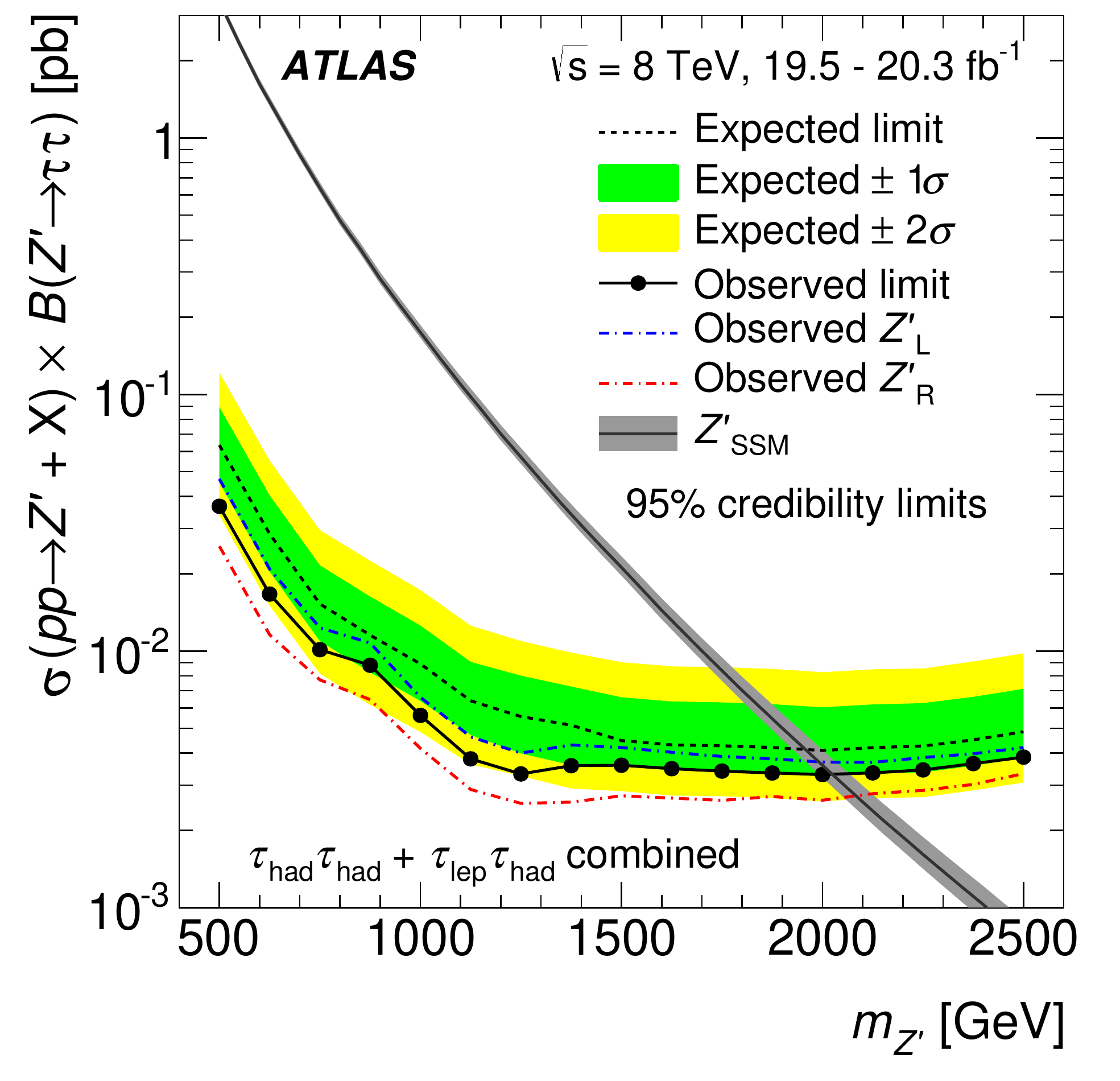}
    \caption{
        Bayesian 95\% credibility upper limits on the cross section
        times ditau branching fraction for a \Zprime{} in the 
        Sequential Standard Model. The figure shows (left) an overlay 
        of the observed (solid lines with filled markers) and
        expected (dashed lines with empty markers) limits 
        in each channel and for the combination, and (right) 
        the combined limit with $1\sigma$ and $2\sigma$ uncertainty bands and  
        an overlay of the impact of the $\ZprimeL$/$\ZprimeR$ 
        models. 
        The width of the \ZprimeSSM{} theory band represents the theoretical uncertainty 
        from the PDF error set, the choice of PDF as well as the strong coupling constant. 
     }
    \label{fig:ssm_limit}
\end{figure}

Bayesian 95\% credibility upper limits are set on \xsecbrssm{} as a function of
$m_{\Zprime}$, using the Bayesian Analysis Toolkit~\cite{BAT}.
Figures~\ref{fig:ssm_limit}(left) and \ref{fig:ssm_limit}(right) show the
limits for the individual channels and for the combination, respectively.  The
resulting 95\% CL lower limit on the mass of a \ZprimeSSM{} decaying to \tptm{}
pairs is $\obslimssm\TeV$, with an expected limit of $\explimssm\TeV$.  The
observed and expected limits in the individual channels are, respectively:
\obslimssmhadhad{} and \explimssmhadhad{}~\TeV{} (\hadhad{}); \obslimssmmuhad{}
and \explimssmmuhad{}~\TeV{} (\muhad{}); and \obslimssmehad{} and
\explimssmehad{}~\TeV{} (\ehad{}).  Alteration of the \Zprime{} couplings can
impact the signal acceptance as described in section~\ref{sec:sig_acc}. These
changes translate linearly to the limits on \xsecbrssm{}. Limits on the
\ZprimeL{} and \ZprimeR{} models are shown in figure~\ref{fig:ssm_limit}
(right).  The impact of the choice of the prior on the signal-strength
parameter is evaluated by also considering the {\em reference
prior}~\cite{RefPriors}.  Use of the reference prior improves the limit on
\xsecbrssm{} by a maximum of 10\%, corresponding to an increase of $20\GeV$ in
the observed mass limit.

\begin{figure}[htbp]
\centering
\includegraphics[width=0.49\textwidth]{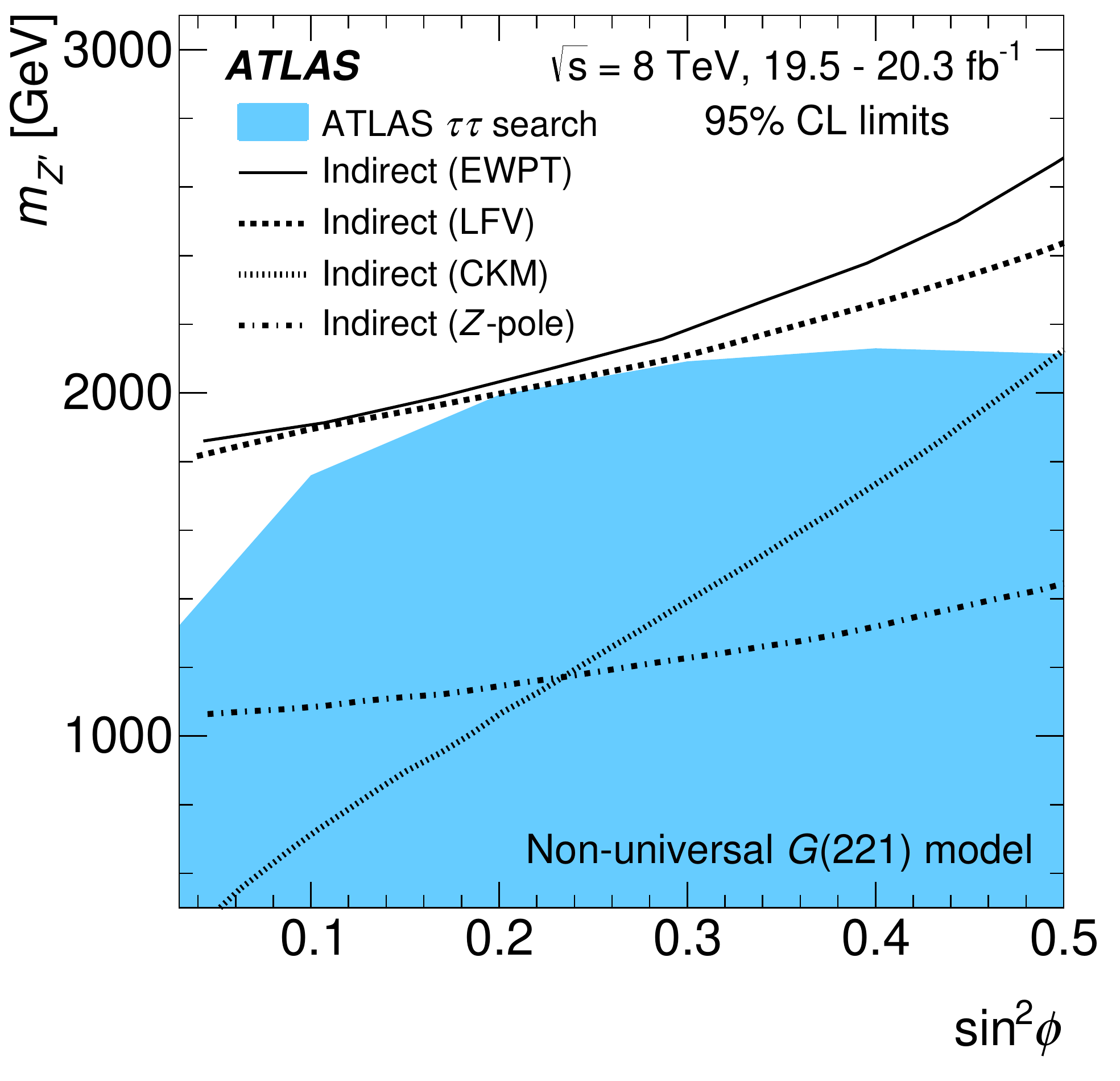}
\caption{ 
  Observed 95\% CL exclusion in the non-universal $G(221)$
  parameter space from the combination of the \hadhad{} and \lephad{} channels
  (blue).  Indirect limits at 95\% CL from fits to electroweak
  precision measurements (EWPT)~\cite{G221LHC}, lepton flavour violation
  (LFV)~\cite{LeeLFV}, CKM unitarity~\cite{LeeCKM} and the original $Z$-pole
  data~\cite{SFM} are overlaid.
}
\label{fig:sfm_limit}
\end{figure}

Limits on the non-universal $G(221)$ model are also calculated. The signal
contributions in the \hadhad{} and \lephad{} channels are rescaled by
$\xsecbrsfm/\xsecbrssm \cdot \acceffsfm/\acceffssm$ (as derived in
section~\ref{sec:sig_sfm}). In addition, the systematic uncertainties are
re-evaluated for each point in parameter space.  Figure~\ref{fig:sfm_limit}
shows the region in the \ZprimeSFM{} parameter space excluded at 95\%
credibility: \ZprimeSFM{} bosons with masses below 1.3--2.1\,TeV are excluded
in the range $0.03<\sin^2\phi<0.5$ assuming no $\mu$--$\tau$ mixing.  Indirect
limits are also overlaid.

%% file: summary.tex
\section{Conclusion}
\label{sec:summary}
A search for high-mass ditau resonances was performed using
19.5--20.3\,$\ifb$ of data collected with the ATLAS detector in $pp$ collisions
at $\sqrt{s}=8\TeV$ at the LHC.  The \hadhad{}, \muhad{} and \ehad{} channels
are analysed.  The observed number of events in regions with high total transverse mass
are consistent with the SM expectations.  Limits are set on the cross
section times branching fraction for such resonances.  The resulting lower
limit on the mass of a \Zprime{} decaying to \tptm{} in the Sequential Standard
Model is $\obslimssm\TeV$ at 95\% credibility, in agreement with the expected
limit of $\explimssm\TeV$ in the absence of a signal. 
The impact on the \Zprime{} acceptance from altering the \Zprime{} couplings
and the total decay width is found to be significant, and can be up to 50\%.
The impact from interference between \Zprime{} and \DYonly{} is only important
when the total decay width of the \Zprime{} is large.
Limits on the non-universal $G(221)$ model are also established, where
\Zprime{} bosons with masses below 1.3--2.1\,TeV are excluded in the range
$0.03<\sin^2\phi<0.5$ assuming no $\mu$-$\tau$ mixing.

%% file: Acknowledgements.tex
\section*{Acknowledgements}


We thank CERN for the very successful operation of the LHC, as well as the
support staff from our institutions without whom ATLAS could not be
operated efficiently.

We acknowledge the support of ANPCyT, Argentina; YerPhI, Armenia; ARC,
Australia; BMWFW and FWF, Austria; ANAS, Azerbaijan; SSTC, Belarus; CNPq and FAPESP,
Brazil; NSERC, NRC and CFI, Canada; CERN; CONICYT, Chile; CAS, MOST and NSFC,
China; COLCIENCIAS, Colombia; MSMT CR, MPO CR and VSC CR, Czech Republic;
DNRF, DNSRC and Lundbeck Foundation, Denmark; EPLANET, ERC and NSRF, European Union;
IN2P3-CNRS, CEA-DSM/IRFU, France; GNSF, Georgia; BMBF, DFG, HGF, MPG and AvH
Foundation, Germany; GSRT and NSRF, Greece; RGC, Hong Kong SAR, China; ISF, MINERVA, GIF, I-CORE and Benoziyo Center, Israel; INFN, Italy; MEXT and JSPS, Japan; CNRST, Morocco; FOM and NWO, Netherlands; BRF and RCN, Norway; MNiSW and NCN, Poland; GRICES and FCT, Portugal; MNE/IFA, Romania; MES of Russia and NRC KI, Russian Federation; JINR; MSTD,
Serbia; MSSR, Slovakia; ARRS and MIZ\v{S}, Slovenia; DST/NRF, South Africa;
MINECO, Spain; SRC and Wallenberg Foundation, Sweden; SER, SNSF and Cantons of
Bern and Geneva, Switzerland; NSC, Taiwan; TAEK, Turkey; STFC, the Royal
Society and Leverhulme Trust, United Kingdom; DOE and NSF, United States of
America.

The crucial computing support from all WLCG partners is acknowledged
gratefully, in particular from CERN and the ATLAS Tier-1 facilities at
TRIUMF (Canada), NDGF (Denmark, Norway, Sweden), CC-IN2P3 (France),
KIT/GridKA (Germany), INFN-CNAF (Italy), NL-T1 (Netherlands), PIC (Spain),
ASGC (Taiwan), RAL (UK) and BNL (USA) and in the Tier-2 facilities
worldwide.

%% file: atlas_authlist.tex
\begin{flushleft}
{\Large The ATLAS Collaboration}

\bigskip

G.~Aad$^{\rm 85}$,
B.~Abbott$^{\rm 113}$,
J.~Abdallah$^{\rm 152}$,
S.~Abdel~Khalek$^{\rm 117}$,
O.~Abdinov$^{\rm 11}$,
R.~Aben$^{\rm 107}$,
B.~Abi$^{\rm 114}$,
M.~Abolins$^{\rm 90}$,
O.S.~AbouZeid$^{\rm 159}$,
H.~Abramowicz$^{\rm 154}$
% ...
\bigskip
\\
% ...
$^{11}$ Institute of Physics, Azerbaijan Academy of Sciences, Baku, Azerbaijan\\
% ...
$^{85}$ CPPM, Aix-Marseille Universit{\'e} and CNRS/IN2P3, Marseille, France\\
% ...
$^{90}$ Department of Physics and Astronomy, Michigan State University, East Lansing MI, United States of America\\
% ...
$^{107}$ Nikhef National Institute for Subatomic Physics and University of Amsterdam, Amsterdam, Netherlands\\
% ...
$^{113}$ Homer L. Dodge Department of Physics and Astronomy, University of Oklahoma, Norman OK, United States of America\\
$^{114}$ Department of Physics, Oklahoma State University, Stillwater OK, United States of America\\
% ...
$^{117}$ LAL, Universit{\'e} Paris-Sud and CNRS/IN2P3, Orsay, France\\
% ...
$^{152}$ Institute of Physics, Academia Sinica, Taipei, Taiwan\\
% ...
$^{154}$ Raymond and Beverly Sackler School of Physics and Astronomy, Tel Aviv University, Tel Aviv, Israel\\
% ...
$^{159}$ Department of Physics, University of Toronto, Toronto ON, Canada\\
% ...
\end{flushleft}
